\documentclass[fleqn,usenatbib]{mnras}

\usepackage{newtxtext,newtxmath}

\usepackage[T1]{fontenc}
\DeclareRobustCommand{\VAN}[3]{#2}
\let\VANthebibliography\thebibliography
\def\thebibliography{\DeclareRobustCommand{\VAN}[3]{##3}\VANthebibliography}

\usepackage{graphicx}	
\usepackage{amsmath}	

\usepackage{subfigure}
\usepackage{color}
\usepackage{multirow}
\usepackage{xcolor}

\title[Reconstruction in $f(Q)$ cosmology]{Data reconstruction of the dynamical connection function in $f(Q)$ cosmology}

\author[Yuhang Yang et al.]{
Yuhang Yang,$^{1,2,3}$ \thanks{E-mail: yyh1024@mail.ustc.edu.cn}
Xin Ren,$^{1,2,3,4}$ \thanks{E-mail: rx76@ustc.edu.cn}
Bo wang,$^{1,2,3}$ \thanks{E-mail: ymwangbo@ustc.edu.cn}
Yi-Fu Cai$^{1,2,3}$\thanks{E-mail: yifucai@ustc.edu.cn}
and Emmanuel N. Saridakis$^{5,2,6}$\thanks{E-mail: msaridak@noa.gr}
\\
$^{1}$Department of Astronomy, School of  Physical Sciences, 
University of Science and Technology of China,  96 Jinzhai Road, Hefei, Anhui 
230026, China\\
$^{2}$CAS Key Laboratory for Research in  Galaxies and Cosmology, 
School of Astronomy and Space Science, \\ \; University of Science and Technology of 
China, 96 Jinzhai Road, Hefei, Anhui 230026, China\\
$^{3}$Deep Space Exploration Laboratory, Hefei 230088, China\\
$^{4}$Department of Physics, Tokyo Institute of Technology, 2-12-1 
Ookayama, Meguro-ku,  Tokyo 152-8551, Japan\\
$^{5}$National Observatory of Athens, Lofos Nymfon, 11852 Athens, 
Greece\\
$^{6}$Departamento de Matem\'{a}ticas, Universidad Cat\'{o}lica del 
Norte, Avda. Angamos  0610, Casilla 1280 Antofagasta, Chile
}

\pubyear{\the\year{}}

\begin{document}
\label{firstpage}
\pagerange{\pageref{firstpage}--\pageref{lastpage}}
\maketitle

\begin{abstract}
We employ Hubble data and  Gaussian Processes   in order to reconstruct the 
dynamical connection function in $f(Q)$  cosmology  beyond the coincident gauge. 
In particular,    there exist three branches of connections that satisfy the 
torsionless and curvatureless conditions, parameterized by   a new dynamical 
function   $\gamma$.  We express the redshift dependence of $\gamma$ in terms of 
the $H(z)$  function and the $f(Q)$ form and parameters, and then we reconstruct 
it using   55 $H(z)$ observation data. Firstly, we investigate the case where 
ordinary conservation law  holds, and we reconstruct the $f(Q)$ function,  which 
is very well described by a quadratic correction on top of Symmetric 
Teleparallel Equivalent of General Relativity. Proceeding to the 
general case,   we consider two of the most studied   $f(Q)$
models of the literature, namely the square-root  and the exponential  one. In 
both cases we reconstruct $\gamma(z)$, and we show that according to AIC and BIC information
criteria its inclusion  is 
favoured  compared to both $\Lambda$CDM paradigm, as well as to the same $f(Q)$ 
models under the coincident gauge. This feature acts as an indication that 
$f(Q)$ cosmology should be studied beyond the coincident gauge.
\end{abstract}




\section{Introduction}

Since the discovery of the acceleration  of the universe expansion in the late 
1990's \citep{SupernovaSearchTeam:1998fmf, SupernovaCosmologyProject:1998vns}, 
the concept of dark energy (DE) was   introduced to explain such an 
unexpected phenomenon. Although  the simplest scenario is just a 
cosmological constant $\Lambda$ \citep{RevModPhys.61.1}, resulting to the 
 $\Lambda$CDM paradigm, the nature of DE remains puzzling. Hence, 
the existence of dark sector along with potential observational cosmological 
tensions  \citep{Planck:2018nkj, Planck:2018vyg, BOSS:2014hwf, 
 Perivolaropoulos:2021jda, Abdalla:2022yfr,DiValentino:2020vvd, Yan:2019gbw,Gangopadhyay:2023nli}, 
opens the way towards  modifications and extensions of the concordance model.

General Relativity (GR) is the standard gravitational theory, and it is based 
on curvature and the   Einstein-Hilbert action \citep{CANTATA:2021ktz}. 
Nevertheless, it is known that 
gravity can be equivalently described through the torsional and  non-metricity 
formulations, namely with Teleparallel Equivalent of 
General Relativity (TEGR) \citep{Aldrovandi:2013wha}   and  
Symmetric Teleparallel Equivalent of General Relativity (STEGR) 
\citep{Nester:1998mp,BeltranJimenez:2017tkd}, respectively.  Together, these 
three
equivalent formulations constitute the geometric trinity of gravity 
\citep{BeltranJimenez:2019esp,Capozziello:2022zzh}. Modifications of curvature-based General 
Relativity  directly lead to the well-known $f(R)$ gravity 
\citep{Starobinsky:1980te, Capozziello:2002rd},  to  $f(G)$
gravity \citep{Nojiri:2005jg},  to Lovelock gravity \citep{Lovelock:1971yv}, 
etc. Furthermore, the 
extension of TEGR, known as $f(T)$ gravity, has been well discussed and studied 
in cosmology \citep{Cai:2015emx,   Cai:2018rzd, Krssak:2018ywd, 
Yan:2019gbw, Huang:2022slc, Wang:2023qfm, Hu:2023juh}.
Finally, modifications based on the non-metricity scalar $Q$, i.e. extensions 
of the STEGR, lead to $f(Q)$ 
gravity \citep{BeltranJimenez:2017tkd,Heisenberg:2023lru}. The cosmological 
applications of   $f(Q)$     gravity   prove to be very interesting, and thus 
they have recently attracted a large amount of research
\citep{Khyllep:2021pcu,Mandal:2020buf,Barros:2020bgg,Jarv:2018bgs,Lu:2019hra,
De:2022wmj,
Solanki:2022rwu,
Lymperis:2022oyo,DAmbrosio:2021zpm,Li:2021mdp,Dimakis:2021gby,
Hohmann:2021ast, 
Kar:2021juu,Wang:2021zaz,Quiros:2021eju,Mandal:2021bpd,Albuquerque:2022eac,
Capozziello:2022wgl,
   Capozziello:2022tvv,Dimakis:2022wkj,DAgostino:2022tdk,
Narawade:2022cgb,
Emtsova:2022uij,Bahamonde:2022cmz,Bahamonde:2022zgj, 
 Sokoliuk:2023ccw, De:2023xua,
Dimakis:2023uib,Maurya:2023szc, Ferreira:2023awf,Capozziello:2023vne,
Koussour:2023rly,Najera:2023wcw,Atayde:2023aoj,
Paliathanasis:2023pqp,Bhar:2023zwi,Mussatayeva:2023aoa,
Paliathanasis:2023kqs,Mandal:2023cag, Pradhan:2023oqo,
Capozziello:2024vix,Bhar:2024vxk,Mhamdi:2024kgu,Goncalves:2024sem}. 
We mention that  recently  there is a 
discussion   whether   $f(Q)$     gravity exhibits the strong 
coupling or      ghost  presence
\citep{Gomes:2023tur}, nevertheless one can construct
versions of the theory that are  free from these issues,  incorporating  non-minimal couplings  or direct couplings of the
matter ﬁeld to the connection   
\citep{Heisenberg:2023lru,Heisenberg:2023wgk,DAmbrosio:2023asf}.

While  the coincident gauge is commonly used in $f(Q)$ cosmology, exploring the 
general covariant formulation provides additional forms of affine connections 
\citep{Zhao:2021zab, Hohmann:2021ast}. Generally, there are three possible 
branches of connections satisfying    
  the torsionless and curvatureless conditions, introducing a 
free 
dynamical function  $\gamma(t)$, which affects the solutions of the theory 
 \citep{DAmbrosio:2021pnd, Heisenberg:2022mbo, Dimakis:2022rkd, Shabani:2023xfn,
Paliathanasis:2023pqp, Dimakis:2022wkj}. One of the three branches is 
equivalent 
to the coincident gauge in cartesian Friedmannn-Robertson-Walker metric 
metric \citep{Hohmann:2021ast}, while the other two   exhibit distinct 
dynamical 
behavior when $\gamma(t)$ is non-vanishing. Therefore, it is of great significance to study the physical properties and evolutionary characteristics of parameter $\gamma(t)$ in $f(Q)$ gravity.

In this work, we are interesting in investigating the dynamical connection
function $\gamma(t)$ of the covariant $f(Q)$ cosmology, from the data 
perspective. There is no physical motivation to determine the functional form of $\gamma(t)$. In particular, we desire to 
 reconstruct it from the data without any assumption of its 
functional form, in contrast to previous studies \citep{Shi:2023kvu, 
Subramaniam:2023okn}. For this sake, we employ Gaussian Processes (GP) for 
data reconstruction, a technique widely used in cosmology, allowing us to 
directly obtain reconstruction functions from observational Hubble parameter 
data \citep{Cai:2019bdh, Seikel_2012, Ren:2021tfi, Ren:2022aeo, 
LeviSaid:2021yat, Bonilla:2021dql, Bernardo:2021qhu, Elizalde:2022rss, 
Yu:2017iju}. 
Subsequently, we analyze the evolution characteristics of the   connection
function and its influence on the dynamics of the universe in 
different branches.

This article is structured as follows.  In Section~\ref{sec:f(Q) gravity and 
cosmology}  we present a concise introduction to covariant $f(Q)$ gravity and 
cosmology. Then, in Section \ref{sec:data and result} we reconstruct the 
  connection function $\gamma(t)$ from the data. In particular, in  
subsection~\ref{sec:data} we display the data list that we use, and we describe 
the Gaussian Process that we apply. Then, in subsection \ref{sec:approach 2} we 
perform the reconstruction procedure assuming the ordinary conservation law for 
the matter sector, while in subsection \ref{sec:approach 1} we present  
the reconstruction results in the general case, considering two specific $f(Q)$ 
models that are the most well studied in the literature.  Finally we draw the 
conclusions and provide some discussion in Section~\ref{sec:conclusion and 
discussion}.

\section{$f(Q)$ gravity and cosmology} \label{sec:f(Q) gravity and cosmology}

In this section we  briefly review $f(Q)$ gravity and its application in 
cosmology. In $f(Q)$ gravity, metric and connection are treated on equal 
footing, necessitating the use of the Palatini formalism to describe 
gravitational interaction \citep{BeltranJimenez:2018vdo}. In this formalism, a 
general affine connection $\Gamma^{\alpha}_{\ \mu\nu}$ is introduced and 
defined 
as
\begin{equation}
 \Gamma^{\alpha}_{\ \mu\nu} = \mathring\Gamma^{\alpha}_{\ \mu\nu}  
+L^{\alpha}_{\ \mu\nu},
\end{equation}
where $\mathring\Gamma^{\alpha}_{\ \mu\nu}$ is the Levi-Civita connection, and 
the $L ^{\alpha}_{\ \mu\nu}$ characterizes the deviation of 
the full affine connection from the Levi-Civita one. In the following, we   use 
the upper ring to represent that the geometric quantity is 
calculated under the Levi-Civita connection. Note that we do  not consider the 
anti-symmetry part 
of 
the connection, since the theory encompasses a torsion-free 
geometry. The affine connection $\Gamma^{\alpha}_{\ \mu\nu}$ establishes the 
affine structure, governing how tensors should be transformed, and defining the 
covariant derivative $\nabla_{\alpha}$.

Utilizing this general affine 
connection, we define the basic object in this theory, the non-metricity 
tensor, as $Q_{\alpha\mu\nu} = \nabla_{\alpha}g_{\mu\nu}$, which characterizes 
the geometry of spacetime. Moreover, the disformation tensor $L^{\alpha}_{\ 
\mu\nu}$ can be expressed  as:
\begin{equation}
    L^{\alpha}_{\ \mu\nu}=\frac{1}{2}(Q^{\alpha}_{\  \mu\nu}-Q^{\ \alpha}_{\mu\ 
\nu}-Q^{\ \alpha}_{\nu\ \mu}).
\end{equation}
By imposing the condition of vanishing curvature, the non-metricity scalar can 
be extracted as
\begin{equation}
     Q=\frac{1}{4}Q^{\alpha}Q_{\alpha}- 
\frac{1}{2}\tilde{Q}^{\alpha}Q_{\alpha}-\frac{1}{4}Q_{\alpha\mu\nu}Q^{
\alpha\mu\nu}+\frac{1}{2}Q_{\alpha\mu\nu}Q^{\nu\mu\alpha},
\label{Qscalar}
\end{equation}
where $Q_{\alpha}=g^{\mu\nu}Q_{\alpha\mu\nu}$ and $\tilde 
Q_{\alpha}=g^{\mu\nu}Q_{\mu\alpha\nu}$  represent the two independent traces of 
the non-metricity tensor. It is convenient   to introduce the 
non-metricity conjugate tensor $P^{\alpha}_{\ \mu\nu}$ as 
\begin{equation}
    P^{\alpha}_{\ \mu\nu}=\frac{1}{4}\left(  -2L^{\alpha}_{\ 
\mu\nu}+Q^{\alpha}g_{\mu\nu}-\tilde{Q}^{\alpha}g_{\mu\nu}-\frac{1}{2}\delta^{
\alpha}_{\mu}Q_{\nu}- \frac{1}{2}\delta^{\alpha}_{\nu}Q_{\mu} \right),
\end{equation}
which allows the non-metricity scalar to be simplified as  
$Q=Q_{\alpha\mu\nu}P^{\alpha\mu\nu}$. 

We can now use the non-metricity scalar $Q$ to construct $f(Q)$ gravity,   
introducing the action
\begin{equation}
    S=\int d^4x\sqrt{-g} \left[ \frac{1}{2}f(Q)+\mathcal{L}_m \right],
    \label{eq:action}
\end{equation}
where $g$ is the determinant of the metric, $f(Q)$  is an arbitrary function of 
$Q$,   $\mathcal{L}_m$  is the 
matter Lagrangian density,  and here we have set the gravitational constant  
$8\pi G=1$.   It is worth noting that STEGR, and therefore GR, is recovered 
for $f(Q) =Q$. Finally, as usual, we define the energy-momentum tensor of matter as
\begin{equation}
    T_{\mu \nu}= 
-\frac{2}{\sqrt{-g}}\frac{\delta(\sqrt{-g}\mathcal{L}_m)}{\delta g^{\mu \nu}}.
\end{equation}

Variation of the above action with respect to the metric leads to the metric 
field equation:
\begin{equation} \label{eq:general metric field equation}
\begin{aligned}
    &T_{\mu\nu}= \\
    &\frac{2}{\sqrt{-g}}\nabla_{\lambda}\left(\sqrt{-g} f_QP^{\lambda}_{\ \mu\nu}\right)-\frac{1}{2}fg_{\mu\nu}+f_Q(P_{\nu\rho\sigma}Q_{\mu}^{\ 
\rho\sigma}-2P_{\rho\sigma\mu}Q^{\rho\sigma}_{\ \ \nu}),
\end{aligned}
\end{equation}
where $f_Q=df/dQ,f_{QQ}=d^2f/dQ^2$. Alternatively,  it can be expressed in a 
covariant formulation to highlight the distinction from GR more clearly, namely
\citep{Subramaniam:2023okn,Beh:2021wva,Zhao:2021zab}:
\begin{equation}
    f_Q \mathring 
G_{\mu\nu}+\frac{1}{2}g_{\mu\nu}(f_QQ-f)+2f_{QQ}P^{\lambda}_{\ 
\mu\nu} \mathring \nabla _{\lambda}Q = T_{\mu\nu},
\end{equation}
where  $\mathring G_{\mu\nu} = \mathring R_{\mu\nu} - \frac{1}{2}g_{\mu\nu} 
\mathring R$ is the standard the Einstein tensor  and
$\mathring \nabla _{\lambda}$ is the  
covariant derivative corresponding to Levi-Civita connection. From 
this formula  we can easily see that in more general cases where $f(Q)$ is not 
a linear function of $Q$, the affine connection will   enter the 
dynamics of the metric, ultimately affecting the solutions.

Variation of action \eqref{eq:action} with respect to the 
connection gives 
\begin{equation}
    4\nabla_{\mu}\nabla_{\nu}(\sqrt{-g}f_QP^{\mu\nu}_{\ \ 
\alpha})=-\nabla_{\mu}\nabla_{\nu}(\sqrt{-g}H_{\alpha}^{\ \mu\nu}),
\end{equation}
where we have introduced the hypermomentum tensor density  
\begin{equation}
    H_{\alpha}^{\ \mu\nu} =-\frac{2}{\sqrt{-g}} \frac{\delta 
(\sqrt{-g}\mathcal{L}_m)}{\delta \Gamma^{\alpha}_{\ \mu\nu}}.
\end{equation}
Finally, using the Bianchi identities, the above connection field 
equation leads to the energy-momentum-hypermomentum conservation law, namely 
\citep{Harko:2018gxr, Hohmann:2021ast,Iosifidis:2021nra,Iosifidis:2020gth}
\begin{equation} \label{eq:general conservarion law}
    \sqrt{-g}\mathring\nabla_{\nu}T_{\mu}^{\ 
\nu}=-\frac{1}{2}  
\mathring\nabla_{\nu}\mathring\nabla_{\rho}(\sqrt{-g}H_{\mu}^{ \ 
\nu\rho}).
\end{equation}

We proceed to the application of $f(Q)$ gravity at a cosmological framework. 
Thus, we consider the isotropic and   homogeneous 
flat  Friedmann-Robertson-Walker (FRW) metric     
\begin{equation}
    ds^2=-dt^2+a(t)^2({dr^2}+r^2d \theta^2+r^2\sin^2\theta d\phi^2),
    \label{eq:FRW metric}
\end{equation}
with $a(t)$ the scale factor. As we mentioned in the Introduction,   most 
studies in $f(Q)$ cosmology were conducted within the coincident gauge choice, 
where all   connection coefficients vanish, namely $\Gamma^{\alpha}_{\ \mu\nu} 
=0$. 
Nevertheless, this is not the only choice  \citep{Hohmann:2021ast}, and 
therefore there are some recent studies which   use different connections in 
$f(Q)$ cosmology 
\citep{Subramaniam:2023okn,Hohmann:2021ast,DAmbrosio:2021pnd,Heisenberg:2022mbo, 
Dimakis:2022rkd,Shi:2023kvu,Paliathanasis:2023pqp,Jarv:2023sbp,Dimakis:2022wkj}. 

 In general,  the nonzero   components of a torsionless connection   in flat 
FRW 
universe are \citep{Hohmann:2019fvf}:
\begin{equation}
\begin{aligned}
    &\Gamma^{t}_{\ tt}=C_1,\quad \Gamma^{t}_{\ rr}=C_2,\quad \Gamma^{t}_{\ 
\theta\theta}=C_2r^2\quad \Gamma^{t}_{\ \phi\phi}=C_2r^2\sin^2{\theta}, \\
    &\Gamma^{r}_{\ tr}=C_3,\quad \Gamma^{r}_{\ rr}=0,\quad \Gamma^{r}_{\ 
\theta\theta}=-r,\quad \Gamma^{r}_{\ \phi\phi}=-r\sin^2{\theta}, \\
    &\Gamma^{\theta}_{\ t\theta}=C_3,\quad \Gamma^{\theta}_{\ 
r\theta}=\frac{1}{r},\quad \Gamma^{\theta}_{\ 
\phi\phi}=-\cos{\theta}\sin{\theta},  \\
    &\Gamma^{\phi}_{\ t\phi}=C_3,\quad \Gamma^{\phi}_{\ 
r\phi}=\frac{1}{r},\quad 
\Gamma^{\phi}_{\ \theta\phi}=\cot{\theta},
\end{aligned}
\end{equation}
where $C_1$, $C_2$, $C_3$ are purely temporal functions. As one can show, in 
total there are three possible branches of such connections that satisfy 
additionally the curvatureless requirement, which are presented in 
 Table~\ref{tab:connection cases}, where 
$\gamma$ is a non-vanishing function on $t$.
\begin{table}
\centering
\caption{Three different branches of  time-dependent functions $C_1, C_2$ and 
$C_3$ 
with vanishing curvature and torsion, where $\gamma(t)$ is a non-vanishing 
dynamical function, and with dots denoting time derivatives.}
\begin{tabular}{l|ccc}
\hline
    Case &$C_1$ &$C_2$ &$C_3$ \\
\hline
    Connection \uppercase\expandafter{\romannumeral1} &$\gamma$ &$0$ &$0$ \\
    Connection \uppercase\expandafter{\romannumeral2} 
&$\gamma+\frac{\dot{\gamma}}{\gamma}$ &$0$ &$\gamma$ \\
    Connection \uppercase\expandafter{\romannumeral3} 
&$-\frac{\dot{\gamma}}{\gamma}$ &$\gamma$ &$0$ \\
\hline
\end{tabular}
\label{tab:connection cases}
\end{table}

In the case of FRW metric, with the above general connection parameterization, 
one can calculate the non-metricity scalar from \eqref{Qscalar} as
\begin{equation}   
Q(t)=3\left[-2H^2+3C_3H+\frac{C_2}{a^2}H-(C_1+C_3)\frac{C_2}{a^2}
+(C_1-C_3)C_3\right],
    \label{eq:Q}
\end{equation}
where $H=\frac{\dot{a}}{a}$ is the Hubble function. Note that when one goes 
back 
to the case of zero connection, the above expression yields the standard result 
$Q(t)=-6H^2 $, while for our three connections of  Table~\ref{tab:connection 
cases} we obtain
\begin{align}
 &Q(t) =-6H^2  &\text{for Connection I } \ ,\\
\label{QconnectionII}
 &Q(t)=3\left(-2H^2+3\gamma H 
+ \dot{\gamma}  \right) &\text{for Connection II }, \\
 &Q(t)=3\left(-2H^2+ \frac{\gamma}{a^2}H+ \frac{\dot{\gamma}}{a^2}
 \right)  &\text{for Connection III} .
 \label{QconnectionIII}
\end{align}

Substituting 
the above   general connection  into the   
field equations \eqref{eq:general metric field equation}, and introducing for 
convenience the ansatz  $f(Q)=Q+F(Q)$, we obtain the modified Friedmann  
equations as
\begin{align}
        \label{eq:Friedmann eqsA}
&\rho_m-\frac{1}{2}F+(\frac{1}{2}Q-3H^2)F_Q-\frac{3}{2
}\dot{Q}(C_3-\frac{C_2 }{a^2})F_{QQ}=3H^2, \\
&p_m+\frac{1}{2}F+(2\dot{H}+3H^2-\frac{1}{2}Q)F_Q-\frac{1}{2}\dot{ Q}(-4H+3C_3+\frac{C_2}{a^2})F_{QQ}   \nonumber \\
&\quad \quad \quad \quad \quad \quad \quad \quad \quad \quad \quad \quad \quad \quad \quad \quad \quad=-2\dot{H}-3H^2, 
      \label{eq:Friedmann eqsB}
\end{align}
where a subscript $Q$ denotes differentiation with respect to $Q$. 
In the above equations, $\rho_m$ and $p_m$ are  the energy density and 
pressure  of the (baryonic plus cold dark matter) matter sector, assuming it to 
correspond to a perfect fluid. Note that we do not consider the radiation 
sector 
since we focus on late-time universe.  Comparing the above Friedmann equations 
with the standard ones, we can see that in the scenario at hand we obtain    an 
effective dark energy sector with energy  density and 
pressure respectively given by
\begin{align}
\rho_{de}&=-\frac{1}{2}F+(\frac{1}{2}Q-3H^2)F_Q-\frac{3}{2}\dot{Q}(C_3-\frac{C_2 }{a^2})F_{QQ},  \\    
p_{de}&=\frac{1}{2}F+(2\dot{H}+3H^2-\frac{1}{2}Q)F_Q-\frac{1}{2}\dot{Q} 
(-4H+3C_3+\frac{C_2}{a^2})F_{QQ}, 
\end{align}
 and thus with  and effective equation-of-state  parameter written as 
\begin{equation}
\begin{aligned}
  w_{de}&=\frac{p_{de}}{\rho_{de}}\\
  &=\frac{\frac{1}{2}F+(2\dot{H}+3H^2-\frac{1}{2} 
Q)F_Q-\frac{1}{2}\dot{Q}(-4H+3C_3+\frac{C_2}{a^2})F_{QQ}}{-\frac{1}{2}F+(\frac{
1 
}{2}Q-3H^2)F_Q-\frac{3}{2}\dot{Q}(C_3-\frac{C_2}{a^2})F_{QQ}}.
\end{aligned}
      \label{eq wde}
\end{equation}

Lastly, let us make some comments on the   the conservation law 
in $f(Q)$ gravity. As one can see from equation \eqref{eq:general conservarion law},
 this relation remains independent of the gravitational part,  depending only 
on 
the matter terms. The left hand side  in the case of FRW metric gives as usual 
   $\mathring\nabla_{\nu}T^{\ 
\nu}_{\mu}=[\dot{\rho}_m+3H(\rho_m+p_m)]u_{\mu}$. In the case of zero 
connection, where the  hypermomentum is absent, the right hand side disappears, 
and thus equation \eqref{eq:general conservarion law} gives 
$\dot{\rho}_m+3H(\rho_m+p_m)=0$ as expected. However, for the general 
connection choices given in Table~\ref{tab:connection cases} 
  the hypermomentum is not always vanish, and thus the dynamical function  
$\gamma(t)$ will enter the right hand side leading to a non-conservation of the 
matter sector. In other words we obtain an effective interaction between the 
connection structure of the geometry and the matter sector, which is typical in 
more complicated geometries 
\citep{Ikeda:2019ckp,Papagiannopoulos:2017whb,
Konitopoulos:2021eav,Savvopoulos:2023qfh}.

Under  Connection \uppercase\expandafter{\romannumeral1} of Table~\ref{tab:connection cases}, the scenario is consistent with the standard 
approach in $f(Q)$ cosmology, where the coincident gauge with vanishing affine 
connections is   used. As a result, the unknown dynamical function 
$\gamma$ 
does not play any role in the evolution of the universe at the background 
level, and this is the aspect which previous 
studies have focused on. However, the situation changes when one considers 
the remaining two types of connections. From the modified Friedmann 
equations  \eqref{eq:Friedmann eqsA},\eqref{eq:Friedmann eqsB}, it is evident 
that $\gamma$ enters   as a new dynamical field, and thus its evolution will   
affect the universe dynamics   at the background level. Furthermore, as 
we 
mentioned above, from equation \eqref{eq:general conservarion law} we deduce 
that   $\gamma$ will play a role in the matter equation, too. We mention here that for different models the evolution feature of $\gamma$ will be different. In the next 
section 
we will investigate 
these particular effects in detail for some specific models, and we will perform a confrontation with 
observational data.

\section{Reconstructing the dynamical connection function $\gamma(t)$ from the 
data}  \label{sec:data and result}

In this  section we desire to   explore  the effect of 
Connection \uppercase\expandafter{\romannumeral2} and Connection
\uppercase\expandafter{\romannumeral3} on the background evolution, and in 
particular to use observational data in order to reconstruct $\gamma(t)$. In 
order to achieve that we will apply  Gaussian Processes, since this procedure 
ensures model-independence and thus it will  
 provide insights into the characteristics of $\gamma$ and its impact on the 
cosmological evolution within the framework of general covariant symmetric 
teleparallel theory.

\subsection{Hubble data set} \label{sec:data}

Since in the following we will apply a reconstruction procedure based on 
$H(z)$, in this work we  will use
 $H(z)$ data from the  observational Hubble data  (OHD)   list, gathered 
from several 
studies \citep{Farooq:2016zwm,Zhang:2016tto,Yu:2017iju,Magana:2017nfs}. The 
$H(z)$ data in this list are primarily derived from cosmic chronometer (CC) and 
radial baryon acoustic oscillations (BAO) observations. Cosmic chronometer 
yields $H(z)$ information by measuring age differences between two galaxies at 
distinct redshifts, evolving independently of any specific model 
\citep{Jimenez:2001gg}. On the other hand, radial BAO observations involve 
pinpointing the BAO peak position in galaxy clustering, relying on the sound 
horizon in the early universe \citep{BOSS:2016wmc,BOSS:2016lpe}. In summary, the 
data list is composed of 31 CC data points and 23 radial BAO data points, as 
documented in \citep{Li:2019nux}. Finally, for    the 
current value of the Hubble function   $H_0$ we adopt the 
most recent SH0ES observation of $73.04\pm 1.04$ $\rm km$ $\rm s^{-1}Mpc^{-1}$ 
\citep{Riess:2021jrx}.

However, the discontinuity in data points with error bars, in general  affects 
the smoothness of the reconstructed functions. To address this issue we employ 
Gaussian Processes in Python (GAPP), and thus we result to a continuous $H(z)$ 
function that best fits our discrete data. GP have been widely used for the 
parameter or function reconstruction in various studies 
\citep{Cai:2019bdh,Seikel_2012,Ren:2021tfi,Ren:2022aeo,LeviSaid:2021yat,
Bonilla:2021dql,Bernardo:2021qhu,Elizalde:2022rss,Yu:2017iju}. Concerning 
the covariance 
function in GAPP, 
for the kernel choice in 
our analysis   we select the exponential form, namely:
\begin{equation}
    k(x,x')=\sigma_f^2e^{-\frac{(x-x')^2}{2l^2}},
\end{equation}
where $\sigma_f$ and $l$ are the hyperparameters. 

\begin{figure}
    \centering
    \includegraphics[width=1\linewidth]{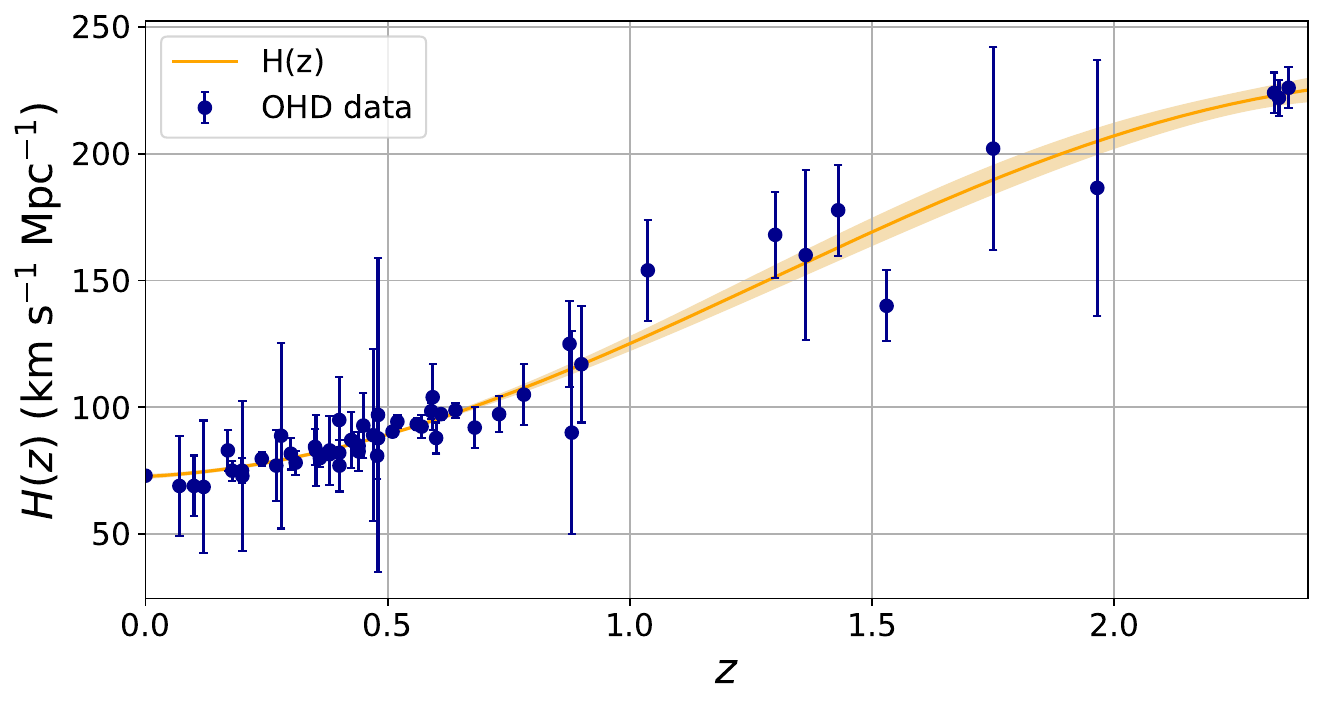}
    \caption{{The reconstructed $H(z)$ function arising from the 55 data 
points through Gaussian Processes,  
imposing $H_0=73.04\pm 1.04$ $\rm km$ $\rm s^{-1}Mpc^{-1}$. The orange curve 
denotes the mean value, while the light yellow shaded zones indicate the 
allowed 
regions at $1\sigma$ confidence level.}}
    \label{fig:H(z) data}
\end{figure}

We apply the GAPP steps, and we obtain a 
  reconstructed $H(z)$ 
function, which is depicted in Figure~\ref{fig:H(z) data}. The orange curve 
represents the mean value, while the light yellow shaded zones indicate the 
allowed regions at   $1\sigma$ confidence level.

\subsection{Reconstruction for the matter conservation 
case}\label{sec:approach 2}

We will start our analysis for the case where there is no interaction between 
the geometry and matter, and thus matter is conserved, namely we will focus on 
the case where the general conservation equation  \eqref{eq:general conservarion 
law} gives  the ordinary conservation law (OCL)
\begin{equation}
 \dot{\rho}_m+3H(\rho_m+p_m)=0.
 \label{OCL}
\end{equation}
Manipulating the Friedmann equations \eqref{eq:Friedmann 
eqsA},\eqref{eq:Friedmann eqsB} we find 
\begin{align}
&\dot{\rho}_m+3H(\rho_m+p_m)  \nonumber \\
&=\dot{Q}\left[-3H^2-\frac{1}{2}Q+ 
\frac{3}{2}\dot{(C_3-\frac{C_2}{a^2})}
+9C_3H-\frac
{3C_2}{a^2}H\right] \nonumber \\  
& \quad +\frac{3}{2}\dot{Q} 
\ddot{Q}\left(C_3-\frac{C_2}{a^2}\right)F_{QQ}+\frac{3}{2}\dot{Q} 
^2\left(C_3-\frac
{C_2}{a^2}\right)F_{QQQ}.
\end{align}

Connection I will lead to trivial results, since $\gamma$ disappears from the 
equations. Let us consider    connection 
\uppercase\expandafter{\romannumeral2} of Table~\ref{tab:connection cases}. In 
this     case, enforcing the OCL 
condition, i.e. enforcing the right-hand-side of the above equation to be 
zero, and using \eqref{QconnectionII}, we obtain
 $(\frac{9}{2}\gamma H \dot{Q} +\frac{3}{2}\ddot{Q}\gamma) 
F_{QQ}+\frac{3}{2}\dot{Q}^2\gamma F_{QQQ}=0$. Thus, applying the chain rules
$\dot{F_Q}= F_{QQ}\dot{Q}$ and 
$\ddot{F_Q}=F_{QQQ}\dot{Q}^2+F_{QQ}\ddot{Q}$, we finally  acquire 
\citep{Shabani:2023nvm}
\begin{equation}
    \ddot{F_Q}+3H\dot{F_Q}=0.
    \label{FQequation}
\end{equation}
In the following it is more convenient to use the redshift $z$ as the 
independent variable, through   $dz/dt=-(1+z)H(z)$. Hence, one can easily 
obtain 
the general solution for   equation \eqref{FQequation} as
\begin{equation}
    F_Q=A\int a^{-3}dt+B = -A\int \frac{(1+z)^2}{H(z)}dz +B,
    \label{eq:conservation fq}
\end{equation}
where $A$ and $B$  are  constants. 
Lastly, note that in the case of Connection III the equations are too 
complicated to accept analytical solutions,  and hence we will not 
consider it further.

Equation \eqref{eq:conservation fq} is our first dynamical equation. The 
second one will arise from the Friedmann equations  \eqref{eq:Friedmann 
eqsA},\eqref{eq:Friedmann eqsB}, which yield
  \begin{equation}
    -2H'\frac{dz}{dt}-\rho_m-p_m=2H' 
F_Q\frac{dz}{dt}-(3\gamma-2H)F_Q^{'}\frac{dz}{dt},
    \label{eq:new fridemann equation}
\end{equation}
where primes denote derivatives with respect to $z$. Finally, the third 
dynamical equation is the OCL \eqref{OCL}.
These three dynamical equations for the three unknown functions, namely $H(z)$, 
$\rho_m(z)$ and $\gamma(z)$ can be easily solved in the case of dust matter, 
namely imposing $p_m=0$. In this case we obtain 
\begin{equation}
    \gamma(z)=\frac{2HH'\left[A \int_0^z  \frac{(1+z')^2}{H(z')}dz' 
-B-1\right]}{3A(1+z)^2}+\frac{2}{3}H(z)+\frac{\Omega_{m0} H_0^2}{A},
    \label{eq:OCL gamma}
\end{equation}
where $\Omega_{m0}$ is the present value of the matter density parameter 
$\Omega_m\equiv \rho_m/(3H^2)$.
Note that in the case   $A=0$  the presence of 
$\gamma$ has no impact on the background evolution, and moreover in such a case 
 \eqref{eq:conservation fq}  leads to a linear  $f(Q)$ form, which is 
consistent 
with our previous discussion that in the standard STEGR case (where $f(Q)$ is 
simply linear in $Q$)   OCL condition holds naturally. In the following we 
focus on the general case where   $A \neq 0$.

Since $F_Q$ is dimensionless, it proves convenient to  parameterize $A$ as 
$A=\epsilon H_0$, with $\epsilon$ a dimensionless parameter and $H_0$ the 
current value of the Hubble function. Additionally, the integration constant 
$B$ in  \eqref{eq:conservation fq} yields a linear term in $f(Q)$, which can 
always be absorbed in a redefinition of the gravitational constant, and hence 
it 
can be set arbitrarily. Without loss of generality, for calculation 
convenience, 
in the following we set it to  $B=-5$. 

Let us now proceed to the reconstruction procedure. In \eqref{eq:OCL gamma} we 
extracted $\gamma(z)$ in terms of $H(z)$, while in the previous subsection, and 
in particular in Figure~\ref{fig:H(z) data}, we    reconstructed $H(z)$ 
from the data. Thus, we can easily reconstruct $\gamma(z)$ itself.
In Figure~\ref{fig:gamma under 
conservation law c2},~\ref{fig:Q under 
conservation law c2} we present the  reconstruction results of $\gamma(z)$ and $Q(z)$
for different values of $\epsilon$. As we observe,   with the 
increase of $\epsilon$ values, the reconstruction 
results  converge to a certain curve, while  the value of 
$\gamma$ at current time is
$8H_0'/3\epsilon+2H_0/3+\Omega_{m0}H_0/\epsilon$. 
Additionally, as the value of   $\epsilon$ increases, the value of $Q$ no longer  decreases monotonically with the expansion of the universe, but instead a minimum value appears in the course of evolution. In the late-time universe, $Q$ appears to evolve independently of the choice of $\epsilon$, which implies that the influence of the existence of $\gamma$ on the universe evolution will diminish progressively.

\begin{figure*}
	\centering
	\subfigure[$\gamma(z)$-$z$ with OCL condition]{
		\begin{minipage}[t]{0.315\linewidth}
			\centering
			\includegraphics[width=\textwidth]{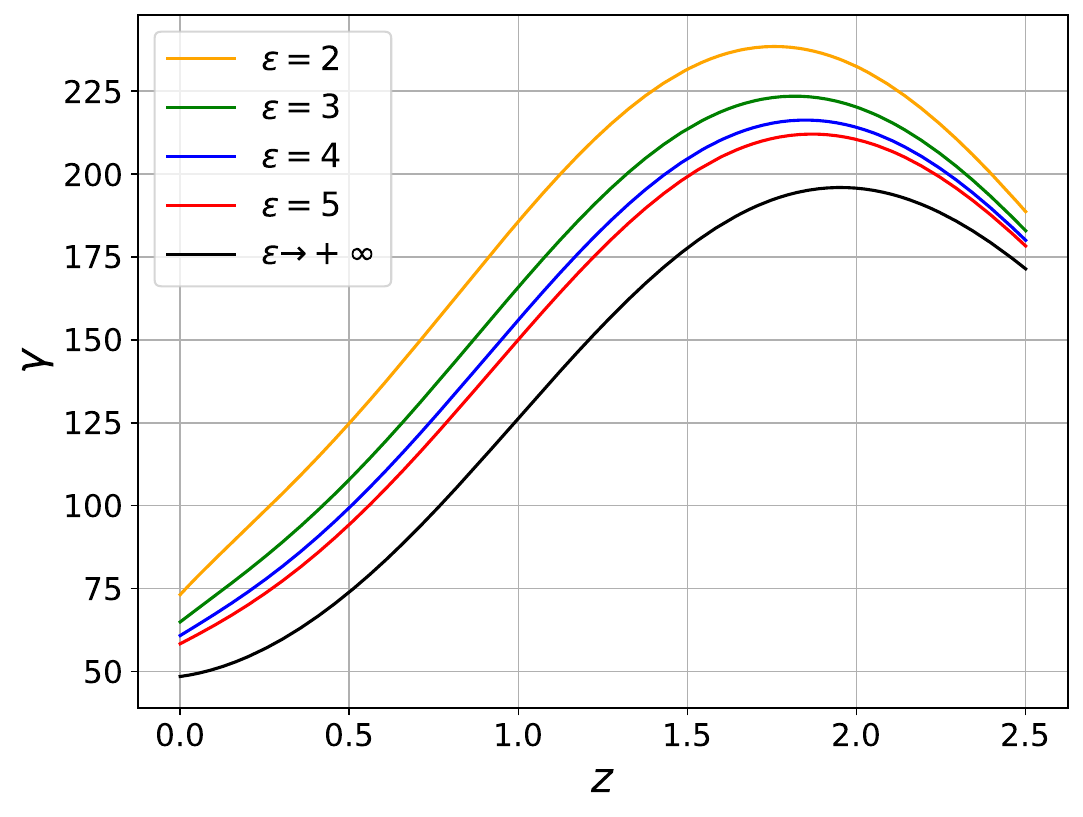}
            \label{fig:gamma under conservation law c2}
		\end{minipage}
	}%
	\subfigure[$Q(z)$-$z$ with OCL condition]{
		\begin{minipage}[t]{0.33\linewidth}
			\centering
			\includegraphics[width=\textwidth]{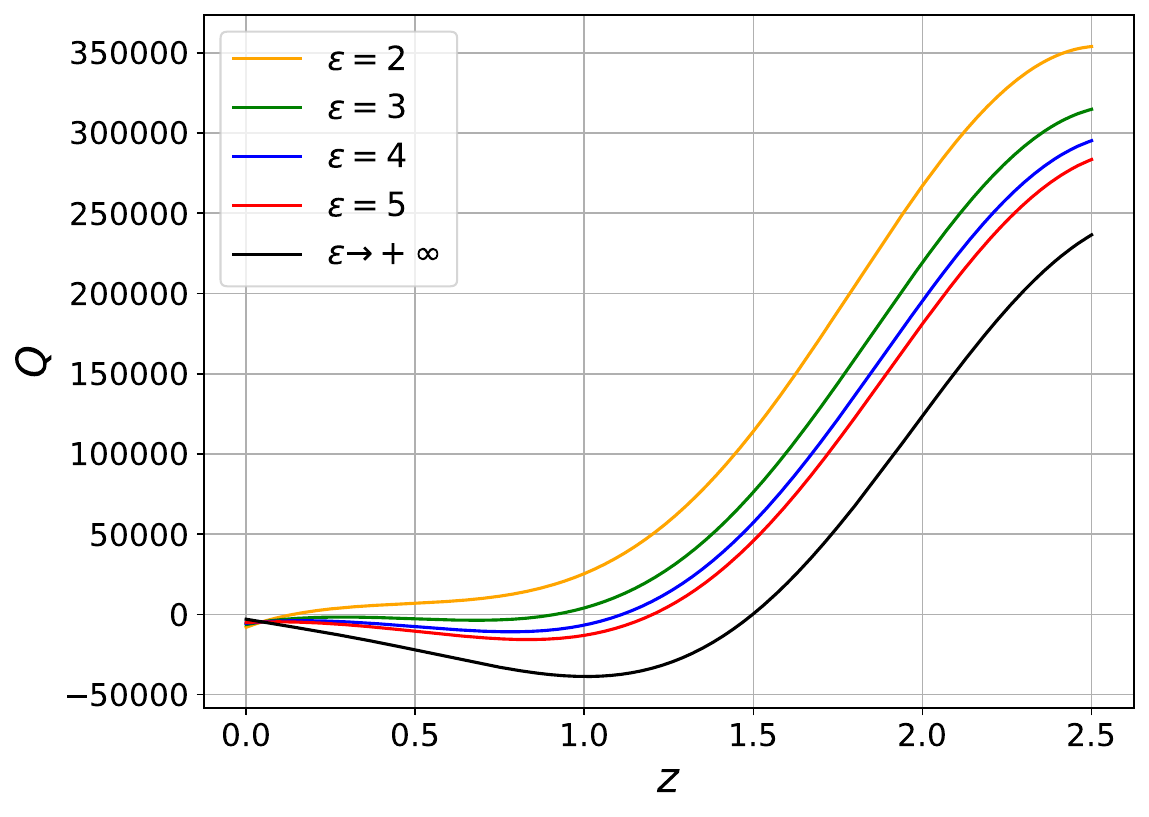}
            \label{fig:Q under conservation law c2}
		\end{minipage}
	}%
	\subfigure[$F(Q)$-$Q$ with OCL condition]{
		\begin{minipage}[t]{0.315\linewidth}
			\centering
			\includegraphics[width=\textwidth]{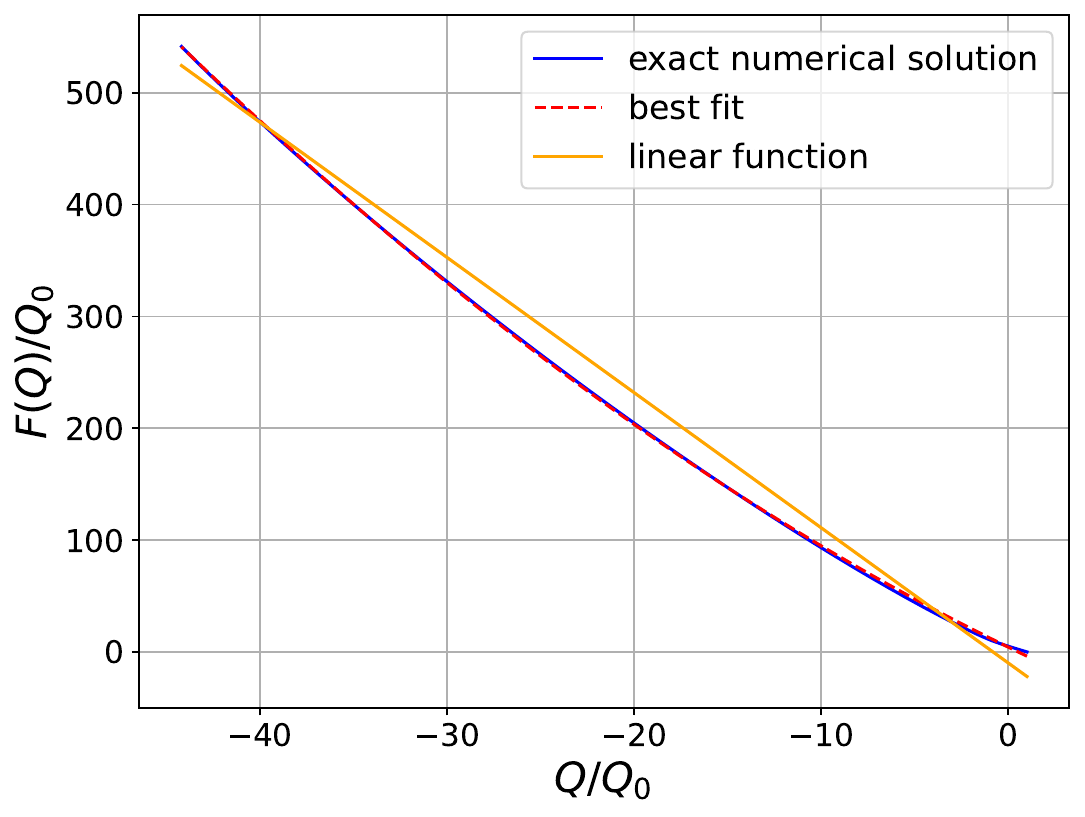}
            \label{fig:F-Q OCL C2}
		\end{minipage}
	}%
	\centering
	\caption{{The reconstructed $\gamma(z)$ and $Q(z)$ for different values of 
$\epsilon$ 
under the assumption of ordinary matter stress-energy tensor conservation law, 
for the case  of Connection \uppercase\expandafter{\romannumeral2}, are shown in Figure~\ref{fig:gamma under conservation law c2} and Figure~\ref{fig:Q under conservation law c2}. The reconstructed $F(Q)$ for 
$\epsilon=2$ are shown in Figure~\ref{fig:F-Q OCL C2}, where the blue solid curve is the exact numerical solution, 
the red dashed curve 
represents the best-fit curve given by $F(Q)=a_1+a_2 Q/Q_0 +a_3 Q^2/Q_0^2$, 
while the orange line depicts the best-fit linear function $F(Q)=b_1+b_2 
Q/Q_0$, namely the STEGR case. The corresponding fitting parameters are 
  $Q_0=-8009$, $a_1=-34614$, $a_2=65317$, $a_3=724$, 
$b_1=78923$, and $b_2=96826$, in $H_0^2$ units. For the reconstruction, we have used the mean values of $H(z)$ presented in Figure~\ref{fig:H(z) data}. Without loss of generality  for numerical calculation we have set $B=-5$.}}
\end{figure*}

In the same lines, using  \eqref{eq:conservation fq} we can reconstruct $ 
F_Q(z)$, while from \eqref{QconnectionII} we can reconstruct $Q(z)$,  thus 
obtaining in the end the reconstruction of  $F(Q)$, which is depicted in  
Figure~\ref{fig:F-Q OCL C2}. Interestingly enough, this function deviates from 
the linear form, hence we deduce that the data favour a deviation from 
standard STEGR, i.e. from standard General Relativity. In particular, taking  
$\epsilon=2$ as an example, we find that   the best fit 
function is the one with a quadratic correction, namely
\begin{equation}
    F(Q)=a_1+a_2 \frac{Q}{Q_0}+a_3 \frac{Q^2}{Q_0^2}, 
    \label{eq:OCL F-Q function}
\end{equation}
with $a_1=-34614$, $a_2=65317$, and $a_3=724$, and where the  current 
value of $Q$ is  $Q_0=-8009$, in $H_0^2$ units. The fact that the quadratic 
correction fits the data very efficiently, and is favoured comparing to 
standard 
General Relativity, is one of the main results of the present work.

\subsection{Reconstruction in the general case} \label{sec:approach 1}

Let us now proceed to the general case, in which the general 
conservation law \eqref{eq:general conservarion 
law} implies an interaction between matter and geometry. In this case it is 
necessary to impose an ansatz for the $f(Q)$ function, and thus in the 
following we will focus on the two most studied models of the literature. 

 The first model is abbreviated 
as Sqrt-$f(Q)$ model, and has the form \citep{Frusciante:2021sio}
\begin{equation}
    F(Q)=M \sqrt{-Q}-2\Lambda,
    \label{eq:Sqrt-f(Q) model}
\end{equation}
where $M,\Lambda$ are free parameters. To render this expression dimensionless, 
we introduce a parameter $\alpha = M/H_0$, yielding the form $F(Q) = (\alpha 
H_0)\sqrt{-Q} - 2\Lambda$. In the coincident gauge, this 
model reproduces the same background evolution with   $\Lambda$CDM 
scenario. However, the 
distinctive effects of varying $\alpha$ become  discernible   by analyzing 
the 
evolution of perturbations in the coincident gauge 
\citep{Barros:2020bgg,Frusciante:2021sio,Atayde:2021pgb,Ferreira:2022jcd}. It 
remains an open question whether these conclusions hold in more general 
connections, as explored in \citep{Subramaniam:2023okn}, where the authors 
investigate the energy conditions under different assumptions on the form of 
$\gamma$. 

The second model  is abbreviated  Exp-$f(Q)$ model, and has the form 
\citep{Anagnostopoulos:2021ydo}
\begin{equation}
    F(Q)=Qe^{\beta \frac{Q_0}{Q}}-Q,
\end{equation}
where $\beta$ is a dimensionless free parameter, and $Q_0$  
the 
value of $Q$ at current time. This model  exhibits a remarkable capability 
to effectively align with observations, and in some cases it is favoured 
comparing to $\Lambda$CDM scenario, although   it does not contain an explicit  
cosmological constant 
\citep{Anagnostopoulos:2021ydo,Khyllep:2022spx,Lymperis:2022oyo,
Ferreira:2023awf}.
  Moreover, it 
effortlessly satisfies Big Bang Nucleosynthesis (BBN) constraints 
\citep{Anagnostopoulos:2022gej}, since at early times, where $Q\gg Q_0$, it 
coincides with General Relativity. Once again we mention that these results     
 stem from analyses conducted in the coincident gauge, prompting further 
investigation into their applicability in more general connections. 

Let us proceed to the reconstruction of the connection function $\gamma(z)$. 
From the  Friedmann equations 
\eqref{eq:Friedmann eqsA},\eqref{eq:Friedmann eqsB} we find that
\begin{equation}
    \dot{Q} 
=\frac{F+2(2\dot{H}+3H^2)(F_Q+1)-QF_Q}{(-4H+3C_3+\frac{C_2}{a^2})F_{QQ}}.
    \label{eq:Dot Q field equation}
\end{equation}
Similarly to the previous subsection, we will not consider Connection I, since 
in this case $\gamma$ does not affect the equations. For the other two 
connections, taking the time derivative of  
\eqref{QconnectionII},\eqref{QconnectionIII} yields:
\begin{align}
\label{eq:Dot Q derive Q0}
&\dot{Q}= 
-12H\dot{H} +9\dot{\gamma}H +9\gamma \dot{H}+3\Ddot{\gamma}  
& \text{for Connection 
\uppercase\expandafter{\romannumeral2} }, \\
&\dot{Q}=-12H\dot{H} +\frac{3\ddot{\gamma}+3\gamma 
\dot{H}-3\dot{\gamma}H-6\gamma 
H^2}{a^2} & \text{for Connection 
\uppercase\expandafter{\romannumeral3}} .
\label{eq:Dot Q derive Q}
\end{align}
Comparing eq.~\eqref{eq:Dot Q field equation} with \eqref{eq:Dot Q derive Q0} 
and \eqref{eq:Dot Q derive 
Q}, leads to
\begin{equation}
    \ddot{\gamma}= 
 \frac{F+2(2\dot{H}+3H^2)(F_Q+1)-QF_Q}{3(-4H+3 \gamma)F_{QQ}} +4H\dot{H} 
-3\dot{\gamma}H -3\gamma \dot{H},
\label{eq: ddot gamma c2}
\end{equation}
for Connection 
\uppercase\expandafter{\romannumeral2} and 
\begin{equation}
    \ddot{\gamma}= a^2\!\left[ 
\frac{F\!+\!2(2\dot{H}\!+\!3H^2)(F_Q\!+\!1)\!-\!QF_Q}{3\left(\frac{\gamma}{a^2}
-4H\right)F_{ QQ }} +4H\dot { H } \right] \!+\!2\gamma H^2\!+\!\dot{\gamma} H 
\!-\!\gamma \dot{H},
\label{eq: ddot gamma c3}
\end{equation}
for Connection 
\uppercase\expandafter{\romannumeral3}.  Hence,  changing the cosmic time $t$ to redshift $z$ through  
$dz/dt=-(1+z)H(z)$, we are now able to reconstruct the evolution  $\gamma(z)$
using the observationally reconstructed  $H(z)$
of Figure~\ref{fig:H(z) data}. It can be seen from  eq.~\eqref{eq: ddot gamma c2} and eq.~\eqref{eq: ddot gamma c3} that the reconstruction of parameter $\gamma(z)$ depends on the specific function form of $f(Q)$.  We mention here that in the literature one can 
find the observational constraints for Sqrt-$f(Q)$ model and Exp-$f(Q)$ model, 
however only for the case of zero connection (coincident gauge). Therefore, 
since in the present work we focus on non-zero $\gamma$, the model parameters 
(namely $\alpha$, $\Lambda$ for the first model and $\beta$ for the second model 
respectively) should   be considered as free parameters from scratch.

\begin{figure*}
	\centering
	\subfigure[Sqrt-$f(Q)$ model]{
		\begin{minipage}[t]{0.32\linewidth}
			\centering
			\includegraphics[width=\textwidth]{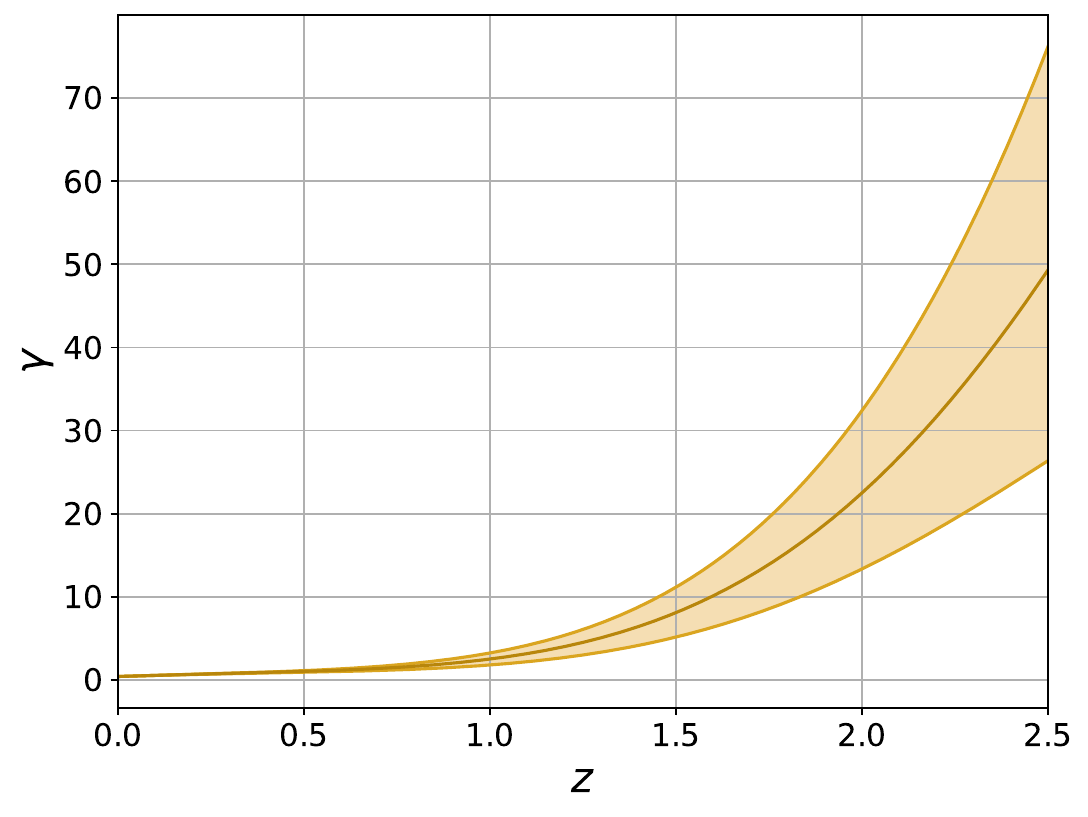}
            \label{fig:gp gamma c2 sqrt l}
		\end{minipage}
	}%
	\subfigure[Sqrt-$f(Q)$ model with $\Lambda=0$]{
		\begin{minipage}[t]{0.325\linewidth}
			\centering
			\includegraphics[width=\textwidth]{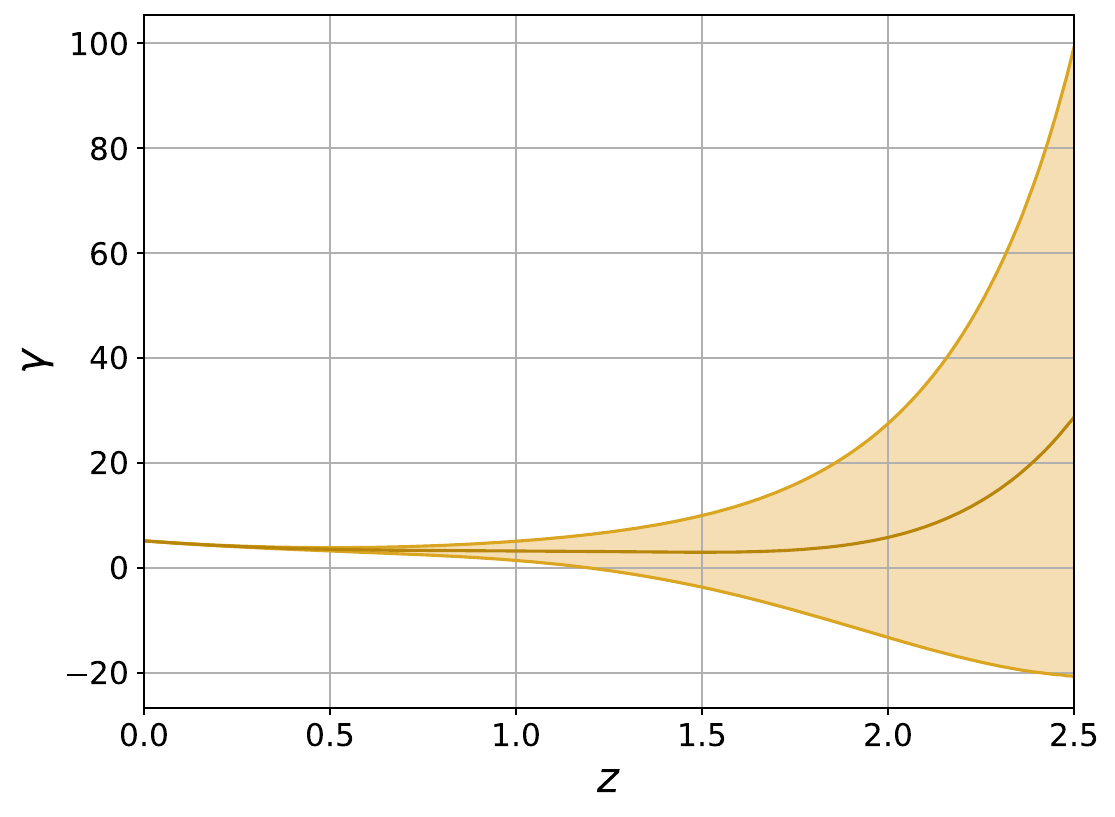}
            \label{fig:gp gamma c2 sqrt no l}
		\end{minipage}
	}%
	\subfigure[ Exp-$f(Q)$ model]{
		\begin{minipage}[t]{0.335\linewidth}
			\centering
			\includegraphics[width=\textwidth]{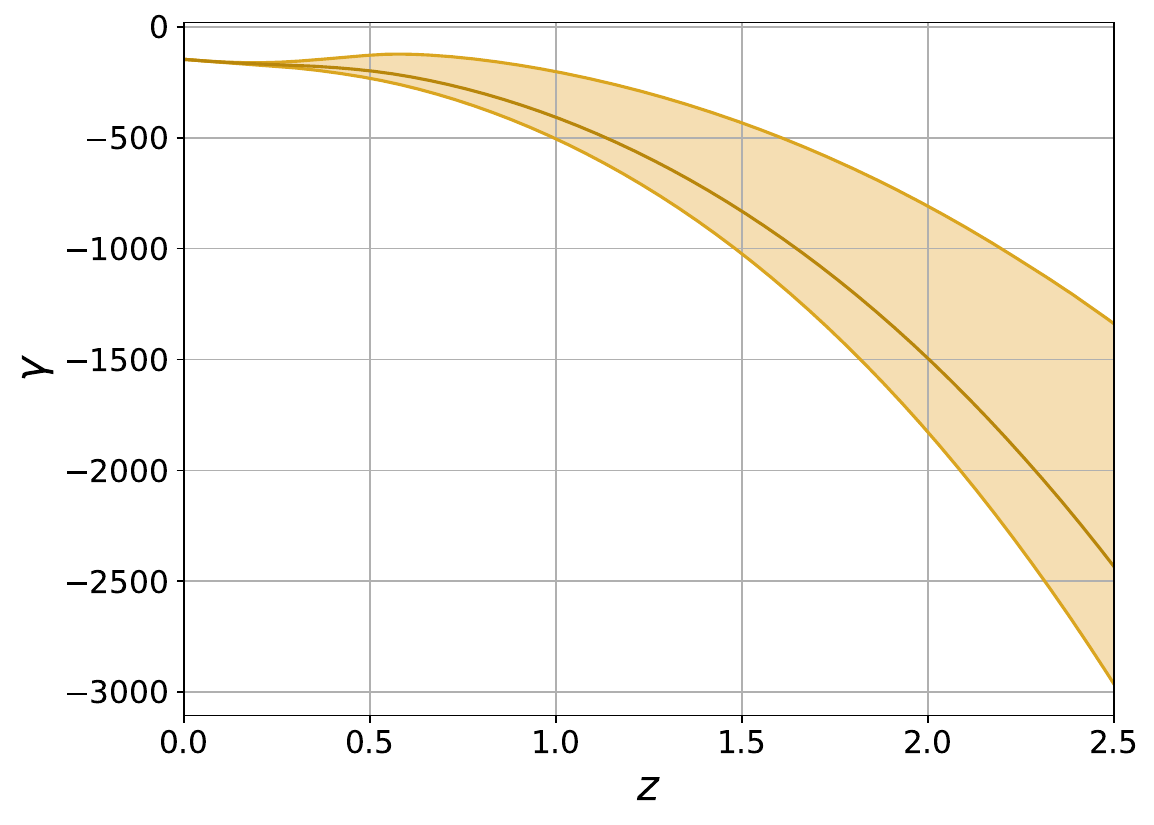}
            \label{fig:gp gamma c2 exp}
		\end{minipage}
	}%
 
	\subfigure[Sqrt-$f(Q)$ model]{
		\begin{minipage}[t]{0.335\linewidth}
			\centering
			\includegraphics[width=\textwidth]
			{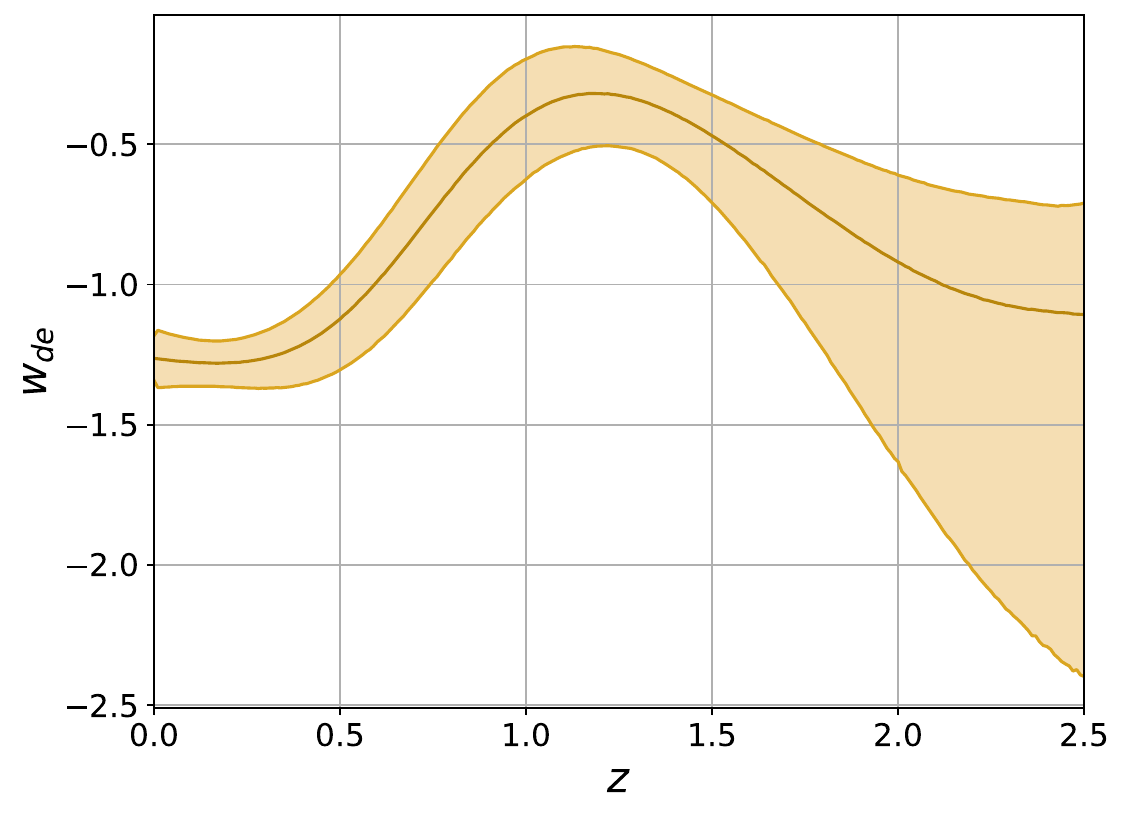}
            \label{fig:gp wde c2 sqrt l}
		\end{minipage}
	}%
    \subfigure[Sqrt-$f(Q)$ model with $\Lambda=0$]{
		\begin{minipage}[t]{0.325\linewidth}
			\centering
			\includegraphics[width=\textwidth]{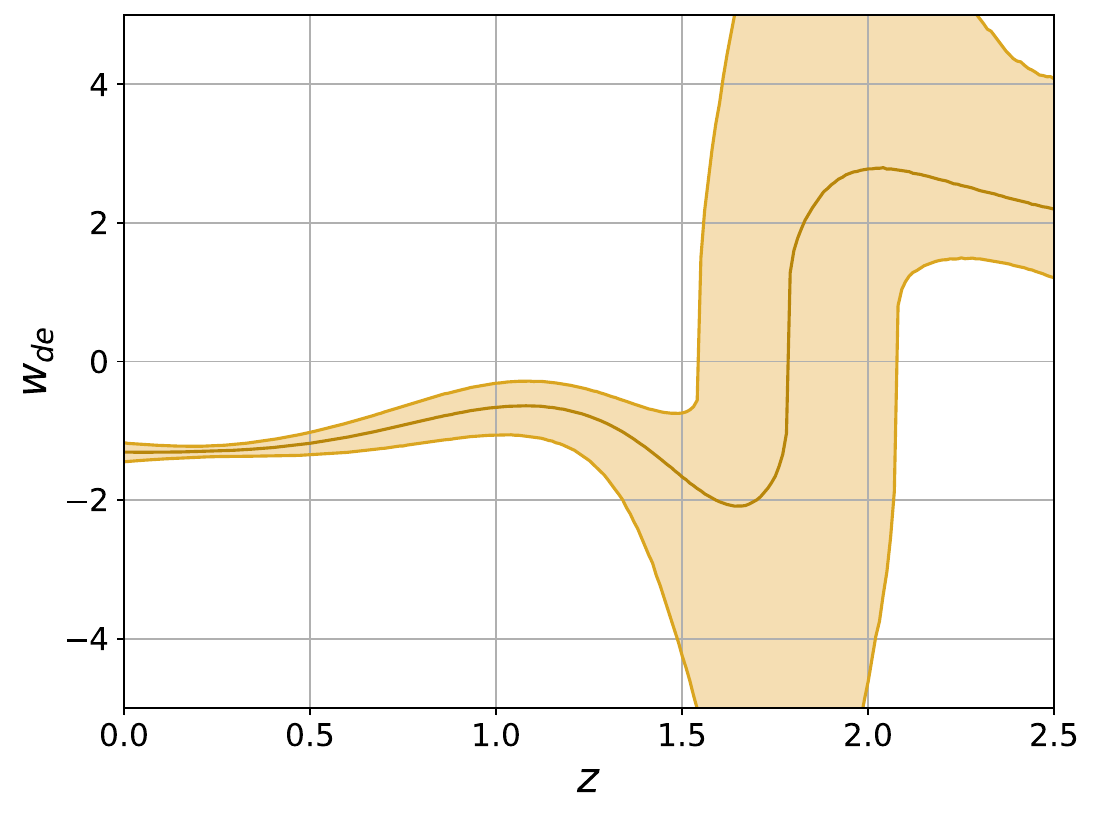}
            \label{fig:gp wde c2 sqrt no l}
		\end{minipage}
	}%
	\subfigure[ Exp-$f(Q)$ model]{
		\begin{minipage}[t]{0.325\linewidth}
			\centering
			\includegraphics[width=\textwidth]{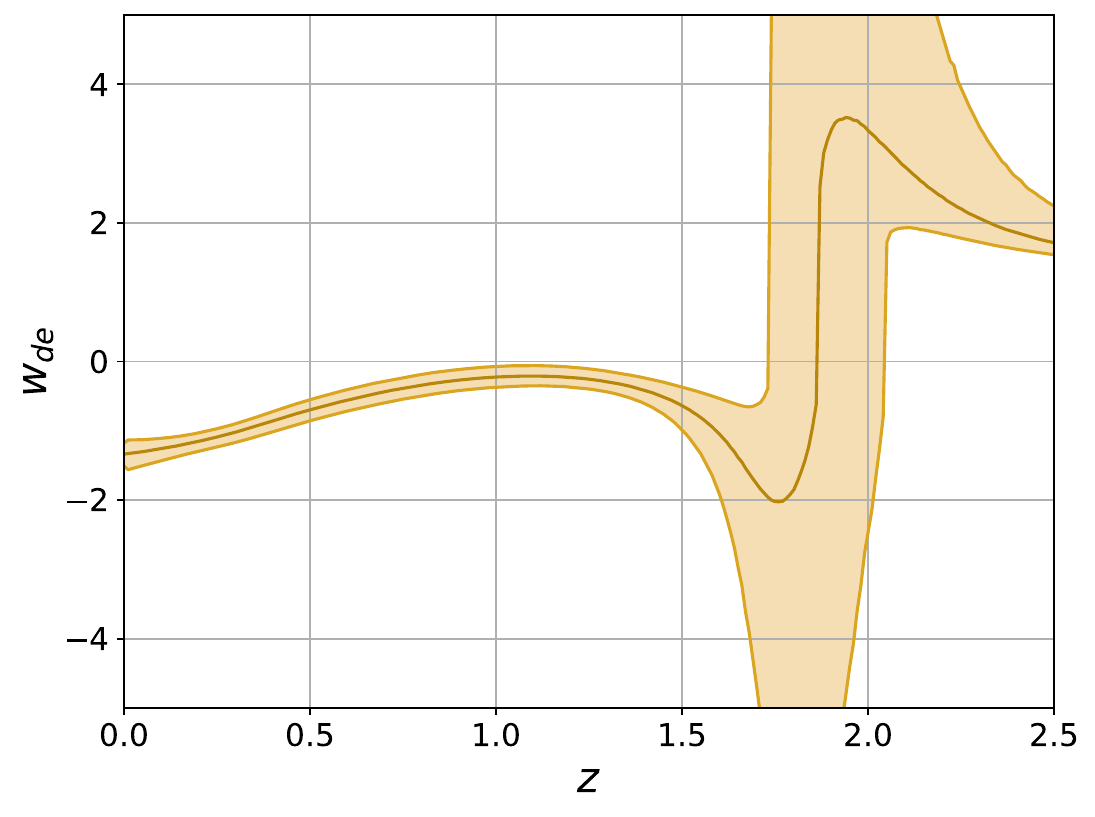}
            \label{fig:gp wde c2 exp}
		\end{minipage}
	}%

    \subfigure[Sqrt-$f(Q)$ model ]{
		\begin{minipage}[t]{0.325\linewidth}
			\centering
			\includegraphics[width=\textwidth]{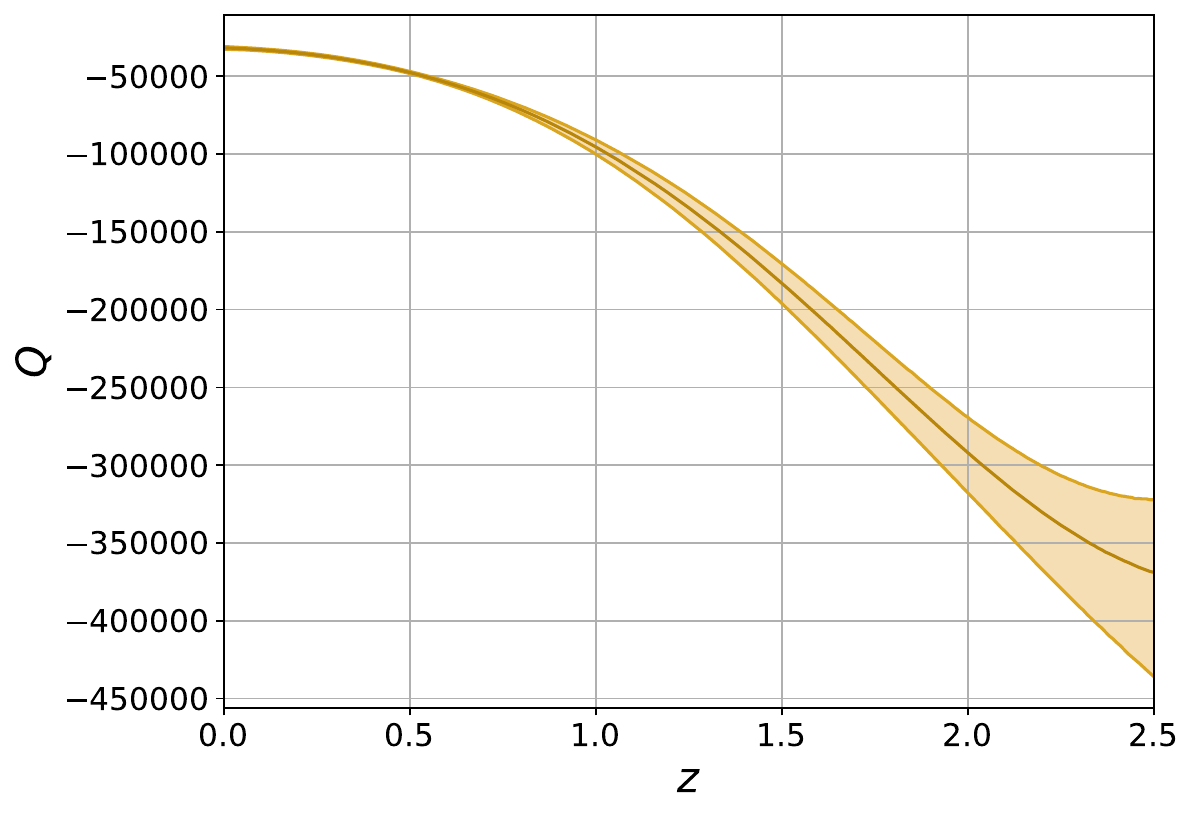}
            \label{fig:gp Q c3 sqrt l}
		\end{minipage}
	}%
    \subfigure[Sqrt-$f(Q)$ model with $\Lambda=0$]{
		\begin{minipage}[t]{0.325\linewidth}
			\centering
			\includegraphics[width=\textwidth]{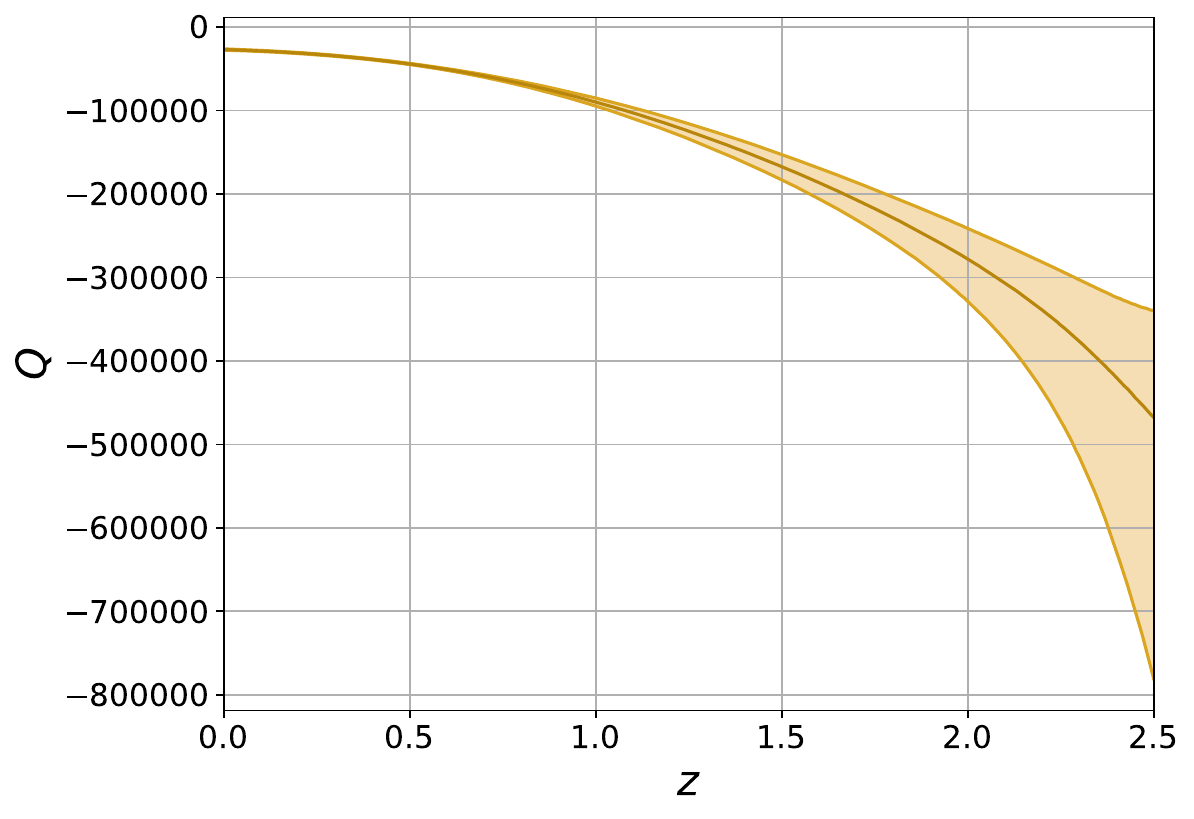}
            \label{fig:gp Q c3 sqrt no l}
		\end{minipage}
	}%
	\subfigure[Exp-$f(Q)$ model]{
		\begin{minipage}[t]{0.325\linewidth}
			\centering
			\includegraphics[width=\textwidth]{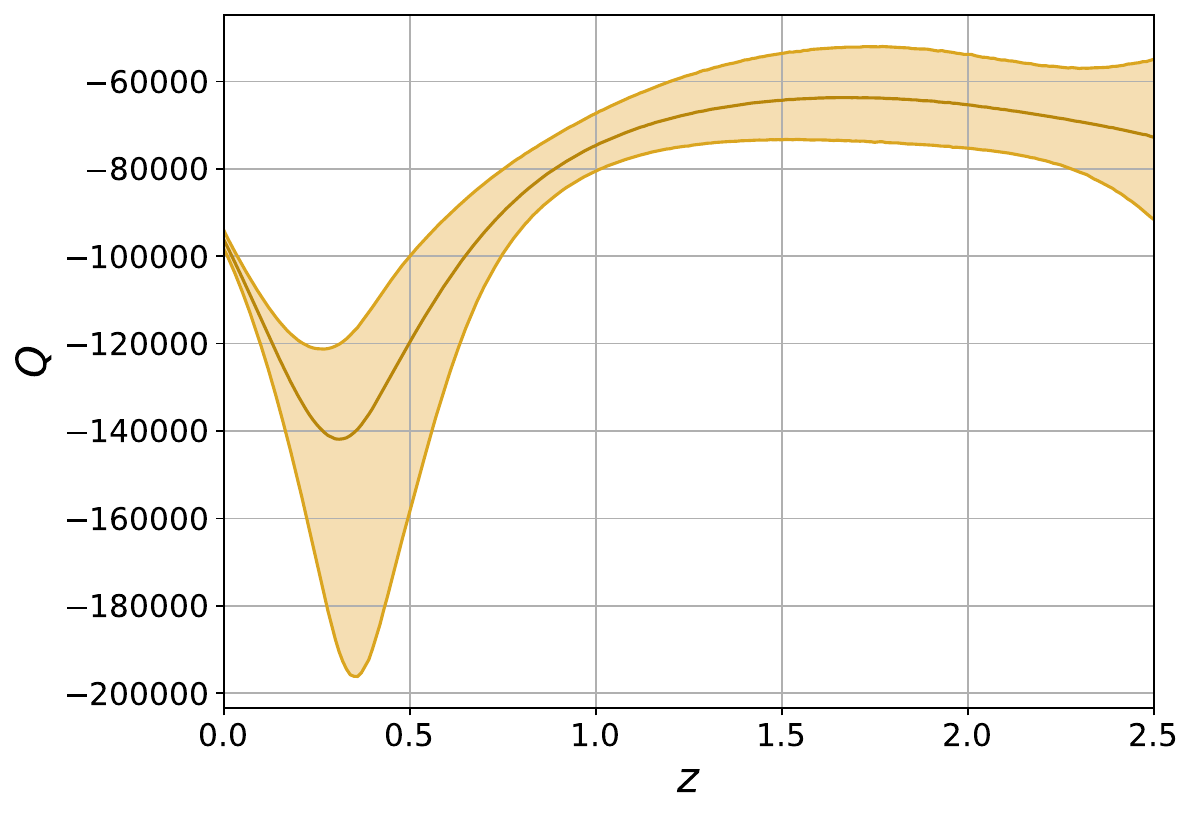}
            \label{fig:gp Q c3 exp}
		\end{minipage}
	}%
	\centering
	\caption{{The reconstructed $\gamma(z)$ (upper panels), the 
corresponding   dark-energy equation-of-state parameter $w_{de}(z)$ (middle 
panels) and non-metricity scalar $Q(z)$ (low panels)}, for the case of 
Connection \uppercase\expandafter{\romannumeral2}. The left panels correspond 
to the  Sqrt-$f(Q)$ model with $M=-1500$ (i.e. $\alpha=-68)$ 
and $\Lambda=0.7\times 3H_0^2$ (the $\Lambda$CDM value), the middle panels 
correspond 
to the  Sqrt-$f(Q)$ model with $M=-5000$ (i.e. $\alpha=-21)$ 
and $\Lambda=0$, while the right panels correspond 
to the  Exp-$f(Q)$ model with $\beta=0.2$, where $M$ is in $H_0$ units.
The bold curves represent the 
mean values, while the shaded areas indicate the $1\sigma$ confidence level.  
}
	\label{fig:GP result connection 2}
\end{figure*}

\begin{figure*}
	\centering
	\subfigure[Sqrt-$f(Q)$ model]{
		\begin{minipage}[t]{0.315\linewidth}
			\centering
			\includegraphics[width=\textwidth]{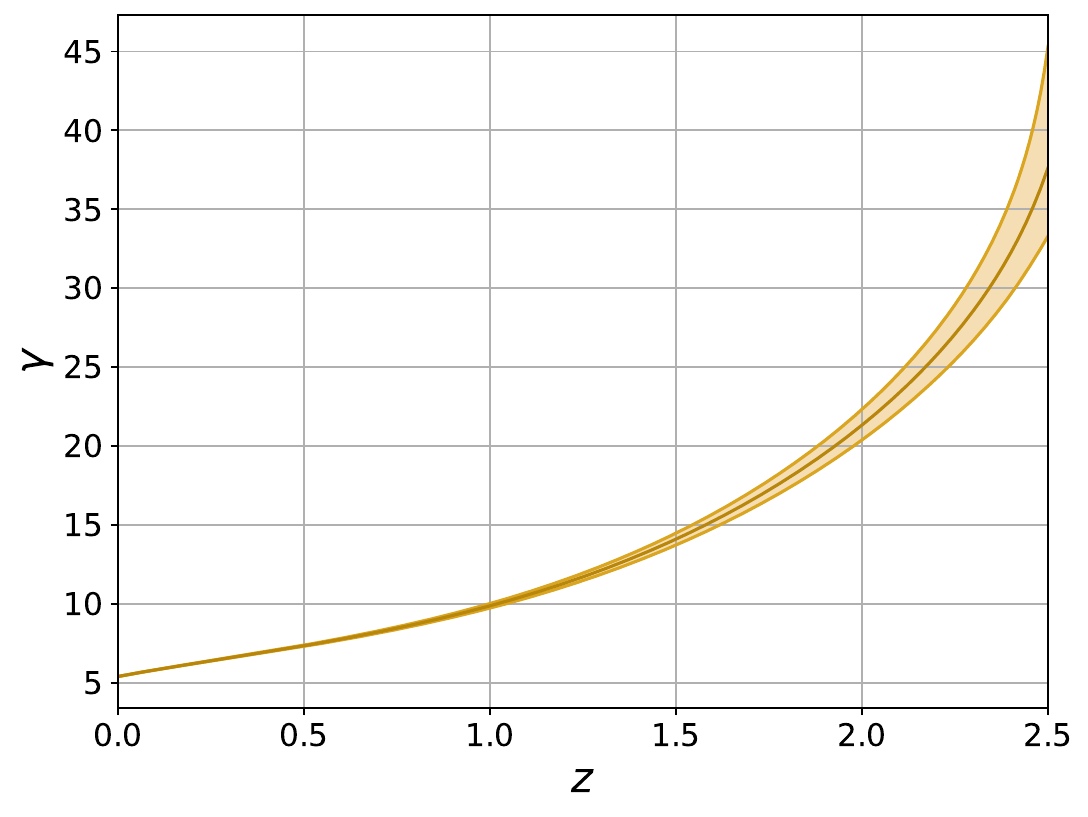}
            \label{fig:gp gamma c3 sqrt l}
		\end{minipage}
	}%
	\subfigure[Sqrt-$f(Q)$ model with $\Lambda=0$]{
		\begin{minipage}[t]{0.325\linewidth}
			\centering
			\includegraphics[width=\textwidth]{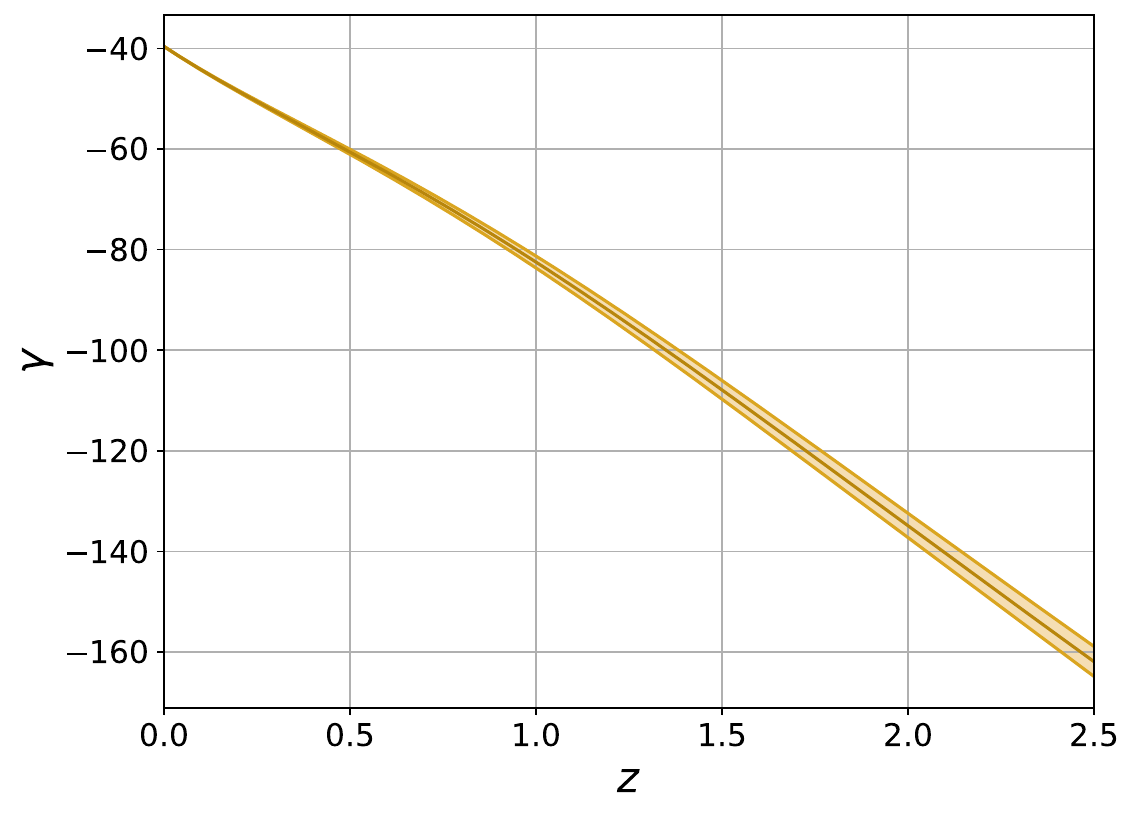}
            \label{fig:gp gamma c3 sqrt no l}
		\end{minipage}
	}%
	\subfigure[Exp-$f(Q)$ model]{
		\begin{minipage}[t]{0.325\linewidth}
			\centering
			\includegraphics[width=\textwidth]{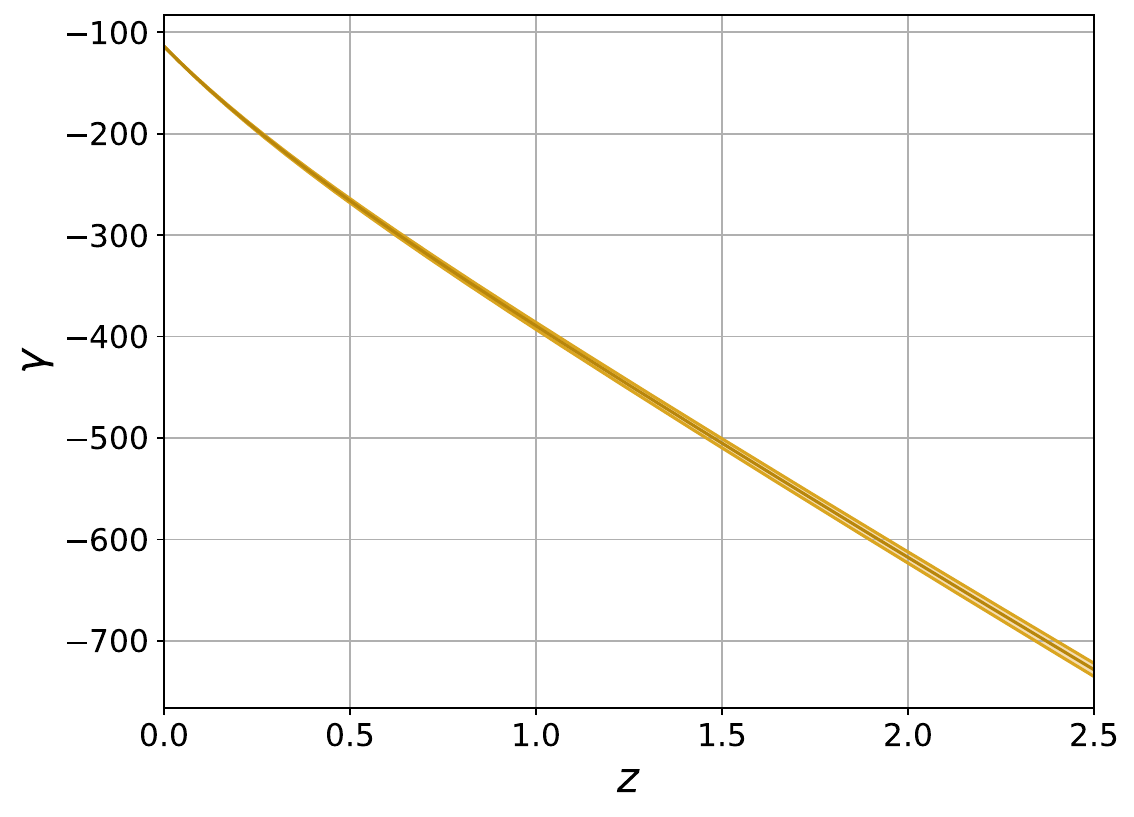}
            \label{fig:gp gamma c3 exp}
		\end{minipage}
	}%
 
	\subfigure[Sqrt-$f(Q)$ model ]{
		\begin{minipage}[t]{0.325\linewidth}
			\centering
			\includegraphics[width=\textwidth]{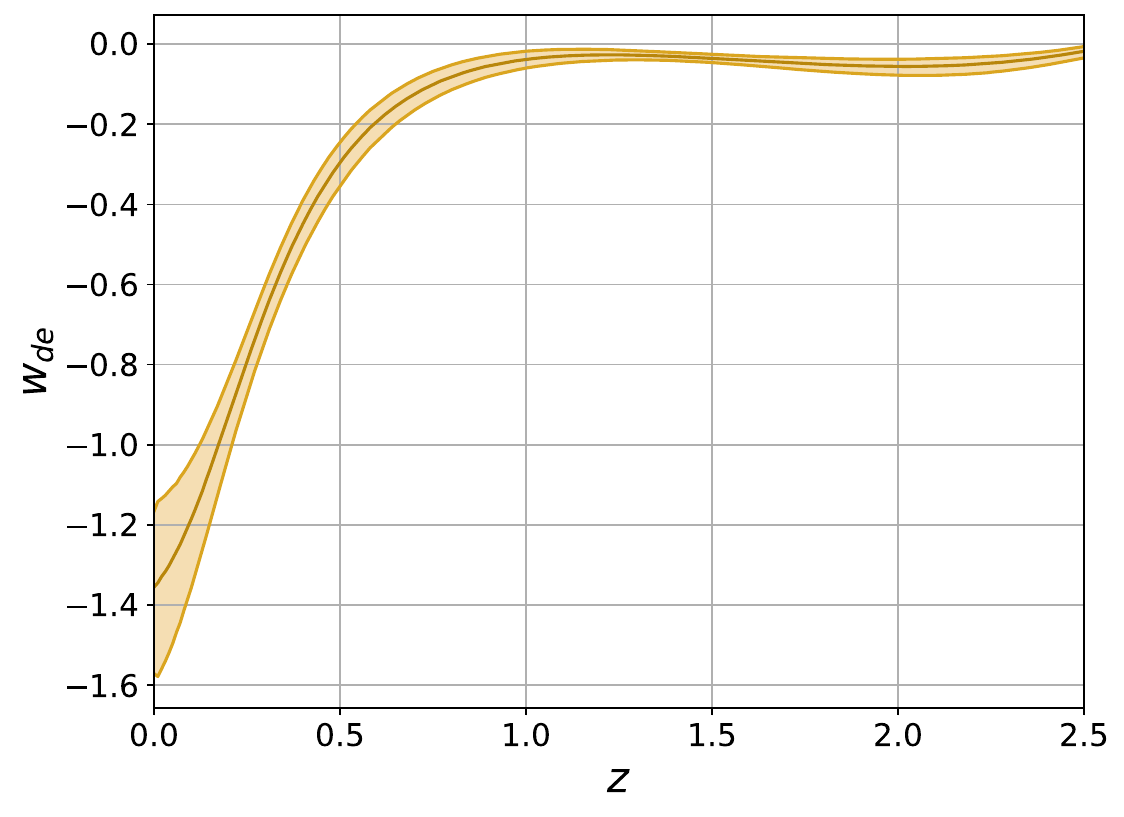}
            \label{fig:gp wde c3 sqrt l}
		\end{minipage}
	}%
    \subfigure[Sqrt-$f(Q)$ model with $\Lambda=0$]{
		\begin{minipage}[t]{0.325\linewidth}
			\centering
			\includegraphics[width=\textwidth]{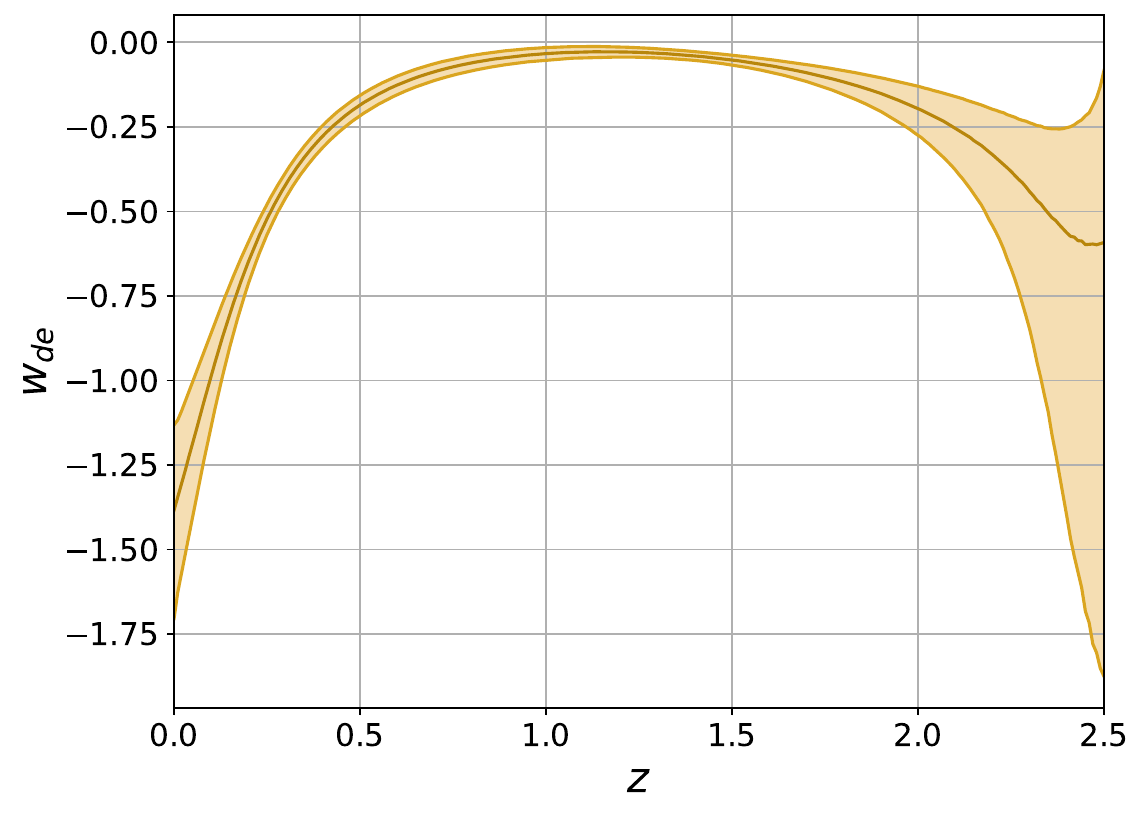}
            \label{fig:gp wde c3 sqrt no l}
		\end{minipage}
	}%
	\subfigure[Exp-$f(Q)$ model]{
		\begin{minipage}[t]{0.325\linewidth}
			\centering
			\includegraphics[width=\textwidth]{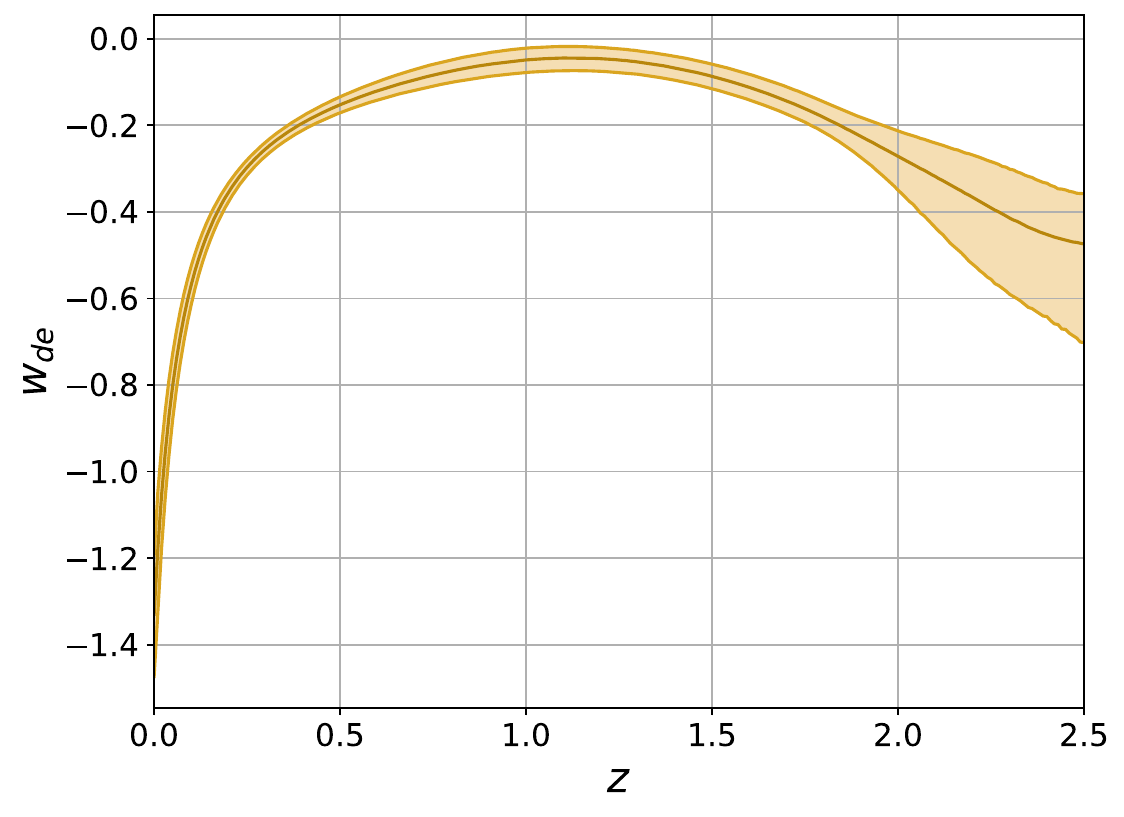}
            \label{fig:gp wde c3 exp}
		\end{minipage}
	}%

    \subfigure[Sqrt-$f(Q)$ model ]{
		\begin{minipage}[t]{0.315\linewidth}
			\centering
			\includegraphics[width=\textwidth]{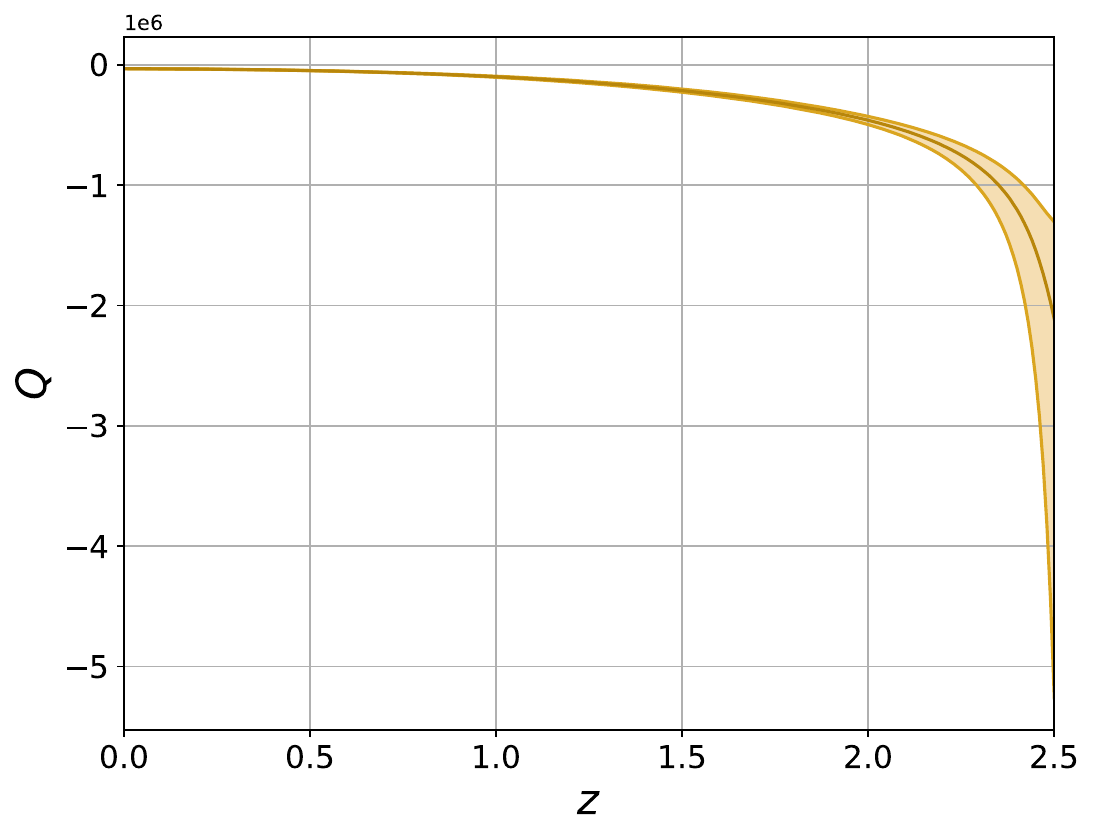}
            \label{fig:gp Q c3 sqrt l}
		\end{minipage}
	}%
    \subfigure[Sqrt-$f(Q)$ model with $\Lambda=0$]{
		\begin{minipage}[t]{0.337\linewidth}
			\centering
			\includegraphics[width=\textwidth]{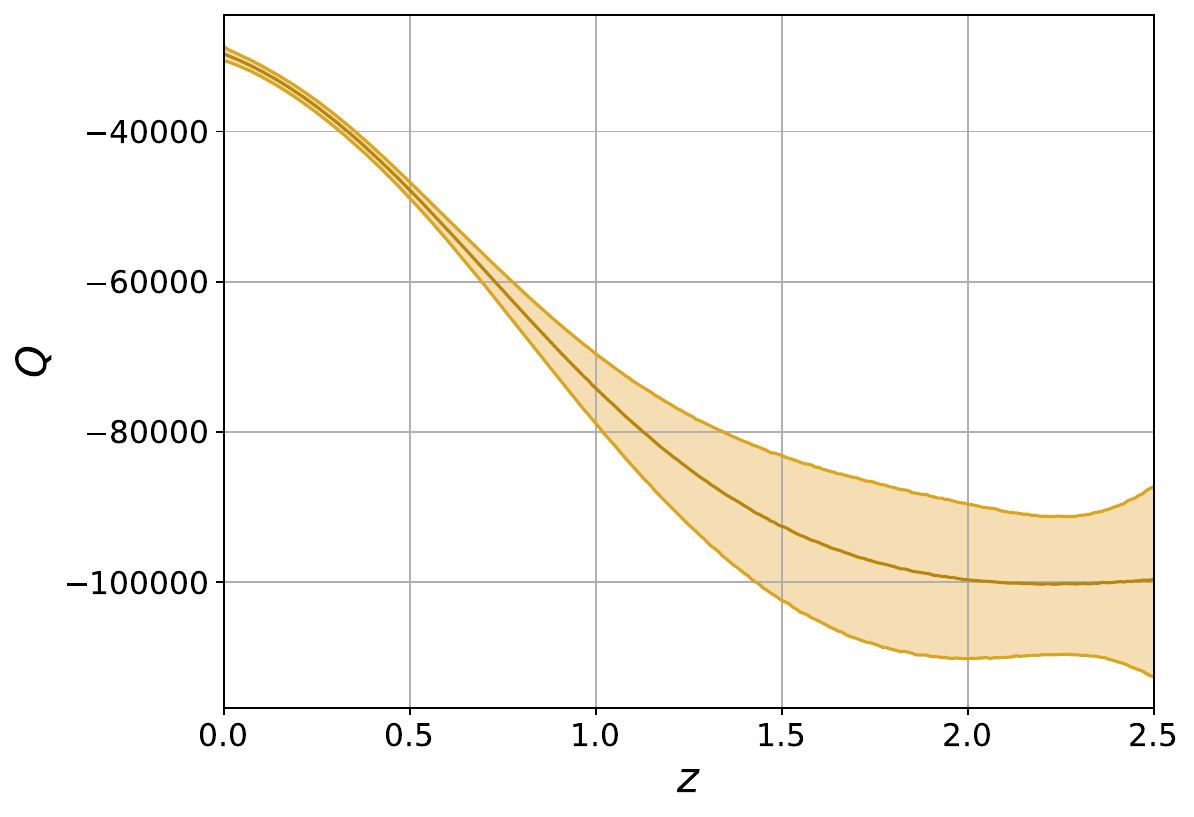}
            \label{fig:gp Q c3 sqrt no l}
		\end{minipage}
	}%
	\subfigure[Exp-$f(Q)$ model]{
		\begin{minipage}[t]{0.325\linewidth}
			\centering
			\includegraphics[width=\textwidth]{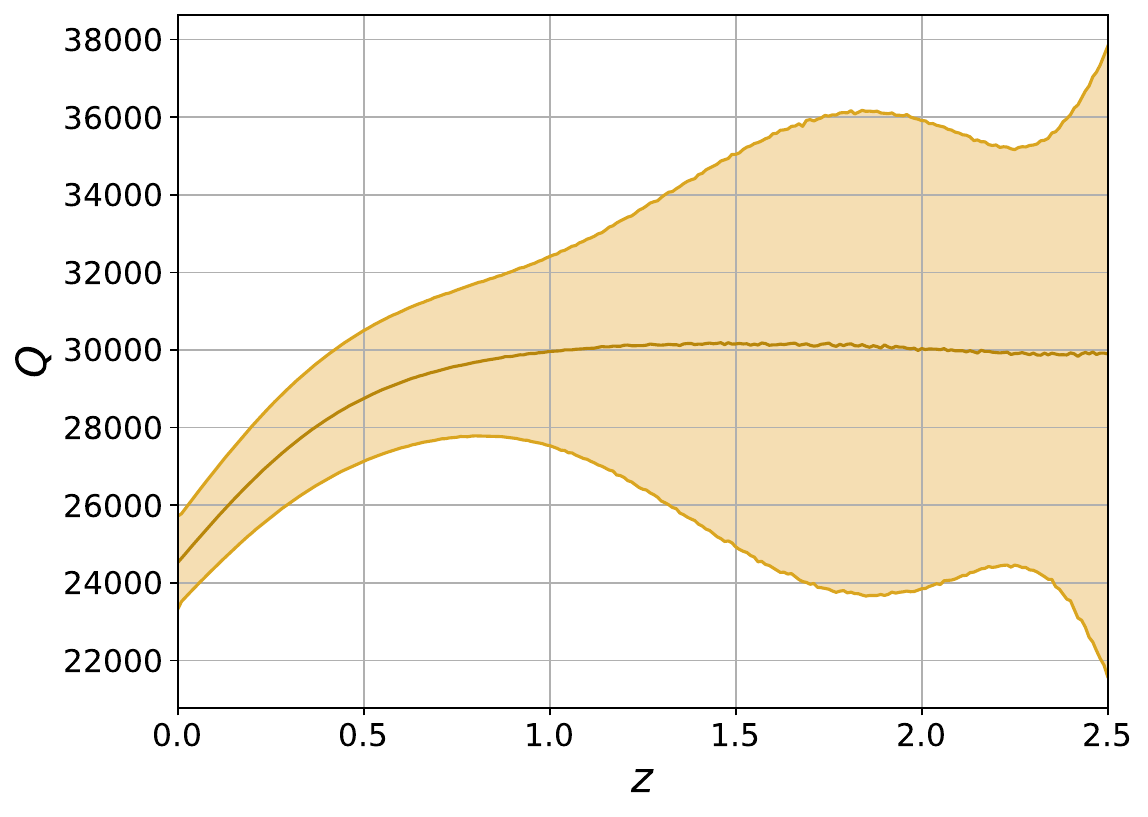}
            \label{fig:gp Q c3 exp}
		\end{minipage}
	}%
	\centering
	\caption{{The reconstructed $\gamma(z)$ (upper panels), the 
corresponding   dark-energy equation-of-state parameter $w_{de}(z)$ (middle 
panels) and non-metricity scalar $Q(z)$ (low panels)}, for the case of 
Connection \uppercase\expandafter{\romannumeral3}. The left panels correspond 
to the  Sqrt-$f(Q)$ model with $M=-1500$ (i.e. $\alpha=-68)$ 
and $\Lambda=0.7\times 3H_0^2$ (the $\Lambda$CDM value), the middle panels 
correspond 
to the  Sqrt-$f(Q)$ model with $M=-5000$ (i.e. $\alpha=-21)$ 
and $\Lambda=0$, while the right panels correspond 
to the  Exp-$f(Q)$ model with $\beta=1.2$, where $M$ is in $H_0$ units.
The bold curves represent the 
mean values, while the shaded areas indicate the $1\sigma$ confidence level.  
}
	\label{fig:GP result connection 3}
\end{figure*}

We start from the case of Connection  \uppercase\expandafter{\romannumeral2}. 
In the upper panels of Figure~\ref{fig:GP result connection 2} we present the 
reconstruction results of $\gamma(z)$, for    Sqrt-$f(Q)$ model with and 
without 
an explicit cosmological constant, and for  Exp-$f(Q)$  model. Additionally, in 
the middle panels of Figure~\ref{fig:GP result connection 2} we depict the 
corresponding reconstructed dark-energy equation-of-state parameter 
$w_{de}(z)$. Finally, in the low panels, the reconstructed non-metricity scalar $Q$ are shown.  Similarly, for Connection \uppercase\expandafter{\romannumeral3}, 
the results are displayed in  Figure~\ref{fig:GP result connection 3}. 
We mention here that in Gaussian Processes  at each specific 
redshift $z$ the value of $H(z)$ exists in the form of a distribution, hence 
there  may be  $H(z)$  values   that can cause divergence in the denominator 
of some terms in the numerical steps, leading to   divergences in the 
reconstructed $w_{de}(z)$ is some cases.

A first and straightforward observation is that the data-driven reconstructed 
$\gamma$ deviate from zero, indicating a deviation from the coincident gauge. 
 Secondly,   as we can see  the reconstruction results have 
significantly smaller errors  in the case 
of connection \uppercase\expandafter{\romannumeral3}.
Thirdly, we mention that although the Sqrt-$f(Q)$ model  in coincident 
gauge exhibits   a degeneracy at the background level   for the different 
model parameter values, and one needs to go at the perturbative level to see 
distinctive effects, in the case of Connections 
\uppercase\expandafter{\romannumeral2} and 
\uppercase\expandafter{\romannumeral3} with a non-zero $\gamma$, different 
$\alpha$ values    lead to different results even at the background level.
Interestingly enough, the Sqrt-$f(Q)$ model can fit the data even in the case 
where $\Lambda=0$, which shows the capabilities in considering non-trivial 
connections. As for non-metricity scalar $Q$, we find that $Q$ evolves differently under different models, but their absolute values will become smaller and smaller in the future as the universe evolves. Finally, concerning  $w_{de}$, as we observe it has a tendency to 
slight phantom values at late times.

\begin{table*}
\caption{The value of $\chi^2$, and the information criteria AIC and BIC  
alongside the corresponding differences $\Delta IC \equiv   IC - IC_{\Lambda 
CDM}$,  for different models. Note that for coincident gauge the Sqrt-$f(Q)$ 
model coincides with $\Lambda$CDM scenario at the background 
level. The Sqrt-$f(Q)$ models with Connection 
\uppercase\expandafter{\romannumeral2} and 
\uppercase\expandafter{\romannumeral3} have been taken without an explicit 
cosmological constant, i.e.   $\Lambda$ =0.}
\centering
\begin{tabular}{c|c|c|c|c|c|c}
\hline 
    \multirow{2}{*}{Model} &\multirow{2}{*}{$\Lambda$CDM} &Coincident gauge
&\multicolumn{2}{|c|}{Connection \uppercase\expandafter{\romannumeral2}} 
&\multicolumn{2}{|c}{Connection \uppercase\expandafter{\romannumeral3}} 
\\
    \cline{3-7}
    & &Exp-$f(Q)$& Sqrt-$f(Q)$ &Exp-$f(Q)$  &Sqrt-$f(Q)$  &Exp-$f(Q)$ \\
\hline
    $\chi^2$ &139.9&51.6 &27.3 &27.2 &26.2 &26.5 \\
    $AIC$ & 141.9& 53.6 & 31.3 &31.2 & 30.2 &30.5 \\
    $BIC$ & 143.9& 55.6& 35.3&35.2&34.2 &34.5 \\
    $\Delta AIC$ &0 &-88.3 &-110.6 &-110.7 &-111.7 &-111.4 \\
    $\Delta BIC$ &0 &-88.3 &-108.6 &-108.7 &-109.7 &-109.4 \\
\hline    
\end{tabular}
\label{tab:chi^2 values}
\end{table*}

We close this section by examining the quality of the fittings. In   
Table~\ref{tab:chi^2 values} we present the $\chi^2$ values, alongside the 
values of the   Akaike Information Criterion (AIC), and the Bayesian Information
Criterion (BIC). The AIC criterion provides an estimator of the 
Kullback-Leibler information and it exhibits the property of asymptotic 
unbiasedness, while and BIC criterion provides an estimator of the Bayesian 
evidence \citep{Liddle:2007fy,Anagnostopoulos:2019miu}. 
Specifically, we have
\begin{eqnarray}
  \chi^2_{model}=\sum_{i=1}^{55}\left [ \frac{(H_{mod,i}-H_{obs,i})^2 }{
\sigma_{H_i}^2}\right]  ,
\end{eqnarray}
while
\begin{eqnarray}
&&AIC=-2\ln{\mathcal{L}_{max}}+2p_{tot}\\
&&BIC=-2\ln{\mathcal{L}_{max}}+p_{tot}\ln{N_{tot}},
\end{eqnarray}
where $H_{mod}$ represents the evolution of the Hubble function corresponding to the specific models with their reconstructed $\gamma$.
And $\ln{\mathcal{L}_{max}}$ represents the maximum likelihood of the model, 
$p_{tot}$ represents the total number of free parameters and $N_{tot}=55$ is the 
number of samples. As we can see, our results indicate that the inclusion of 
$\gamma$ improves the 
quality of the  fittings to observations, compared to both $\Lambda$CDM 
paradigm, as well as to Sqrt-$f(Q)$ and Exp-$f(Q)$ models under the coincident 
gauge. This is one of the main results of the present work. Lastly, for 
completeness we mention that  Connection 
\uppercase\expandafter{\romannumeral3} seems to confront with the data in a 
slightly better way than Connection
\uppercase\expandafter{\romannumeral2}.

\section{Discussion and conclusions } 
\label{sec:conclusion and discussion}

Recently,      $f(Q)$ gravity has garnered significant attention and 
has been the subject of extensive research, since its cosmological applications 
proves to be very interesting. Although the theory has been confronted with 
observations in order to extract information on the possible forms of the 
unknown function $f(Q)$,  almost   all the corresponding analyses have been 
performed under  the coincident gauge. Hence, investigating $f(Q)$ cosmology 
under different connection choices is a subject both interesting and necessary.
In particular, since for general connections that satisfy the torsionless and 
curvatureless conditions a new dynamical function appears, namely $\gamma$, one 
should study the physical implications and evolutionary characteristics of  
$\gamma$, and try to reconstruct it from the observational data themselves. 
 
In this work we used 55 $H(z)$ observation data   to reconstruct the evolution 
of the   dynamical function $\gamma(z)$ of different connections. In 
particular, we first applied  Gaussian  Processes in order to reconstruct 
$H(z)$, and then we expressed  $\gamma(z)$ in terms of  $H(z)$ and the $f(Q)$ 
form and parameters. We studied three different connections beyond the 
coincident gauge, and  we were able to reconstruct  $\gamma(z)$ for 
various cases.

Since   in $f(Q)$ cosmology in general one obtains an effective interaction 
between   geometry and matter, we started our analysis from the case where  
 the  ordinary conservation law  holds. In this case we extracted a general 
solution for the derivative of $f(Q)$, and  utilizing this solution  we 
successfully reconstructed the redshift dependence of $\gamma(z)$, revealing a 
convergence tendency in terms of the model parameter. Additionally,  we 
reconstructed the corresponding $f(Q)$ function,  which is very well described 
by a quadratic correction on top of Symmetric Teleparallel Equivalent of 
General 
Relativity (STEGR).

Proceeding to the general case,   we considered two of the most studied  $f(Q)$ 
models of the literature, namely the square-root (Sqrt-$f(Q)$) one and the 
exponential (Exp-$f(Q)$) one. In both cases we reconstructed $\gamma(z)$, and 
as we showed, the data reveal that $\gamma(z)$ is not zero. However, the most 
interesting result is that the quality of the fitting  after the inclusion of 
$\gamma$ is improved,  compared to both $\Lambda$CDM paradigm, as well as to 
Sqrt-$f(Q)$ and Exp-$f(Q)$ models under the coincident gauge. This feature acts 
as an indication that $f(Q)$ cosmology should be studied beyond the coincident 
gauge.

It would be interesting to confront $f(Q)$ cosmology with non-trivial 
connections with  different observational datasets, such as Supernovae type Ia 
(SN Ia), Cosmic Microwave Background (CMB),   
Redshift Space Distortion (RSD) etc, and moreover examine the 
perturbation evolution. Additionally, it would be necessary to investigate the  non-minimal couplings  and the direct      
connection-matter   couplings, which are able to make the theories free from pathologies. Such analysis lies beyond the scope of 
the present work and it is left for a future project.

\section*{Acknowledgements}
We are grateful to Bichu Li,  Dongdong Zhang, Taotao Qiu, Qingqing Wang, 
Hongsheng Zhao and Jiaming Shi for helpful discussions and valuable comments. 
This work is supported in part by National Key R\&D Program of China 
(2021YFC2203100), by CAS Young Interdisciplinary Innovation Team 
(JCTD-2022-20), 
by NSFC (12261131497), by 111 Project (B23042), by Fundamental Research Funds 
for Central Universities, by CSC Innovation Talent Funds, by USTC Fellowship 
for 
International Cooperation, by USTC Research Funds of the Double First-Class 
Initiative.
We acknowledge the use of computing facilities of TIT, as well as the clusters 
LINDA and JUDY of the particle cosmology group at USTC.

\section*{Data Availability}
The data underlying this article will be shared on reasonable request to the corresponding author.



\bibliographystyle{mnras}
\bibliography{ref}


\bsp	
\label{lastpage}
\end{document}